\documentclass[footinbib,a4paper,aps,pra,reprint,twocolumn,preprintnumbers,amsmath,amssymb,10pt, superscriptaddress]{revtex4-2}
\usepackage{verbatim}
\usepackage{booktabs}
\usepackage{graphicx}
\usepackage{color}
\usepackage{tikz}
\usepackage[colorlinks=true,citecolor=blue,linkcolor=blue,urlcolor=blue]{hyperref}
\usepackage[capitalize]{cleveref} 
\usepackage{braket}
\usepackage{lipsum}  
\usepackage[T1]{fontenc}
\usepackage[normalem]{ulem}
\usepackage{upgreek}
\usepackage[percent]{overpic}
\usepackage{graphicx,xcolor}
\usepackage{dcolumn}
\usepackage{bm}
\usepackage[shortlabels]{enumitem}

\makeatletter
\renewcommand{\fnum@figure}{FIG. \thefigure}
\makeatother

\begin{document}

\title{Critical bifurcation and deconfined quantum criticality \\ in an interacting cluster Ising chain}

\author{Sourabh}
\affiliation{Department of Physics, Indian Institute of Technology Madras, Chennai 600036, India}
\affiliation{Center for Quantum Information, Communication and Computation (CQuICC), Indian Institute of Technology Madras, Chennai 600036, India}

\author{Bachana Beradze}
\affiliation{Department of Engineering and Physics, Karlstad University, Karlstad, Sweden}

\author{Mikheil Tsitsishvili}
\affiliation{Institut für Theoretische Physik, Heinrich-Heine-Universität, D-40225 Düsseldorf, Germany}

\author{Alexander Nersesyan}
\affiliation{Ilia State University, Cholokashvili Avenue 3-5, 0162 Tbilisi, Georgia }
\affiliation{Andronikashvili Institute of Physics, Tamarashvili str. 6, 0177 Tbilisi, Georgia} 
\affiliation{The Abdus Salam International Centre for Theoretical Physics (ICTP), Strada Costiera 11, 34151 Trieste,
Italy}

\author{Titas Chanda}
\email{titas.chanda@physics.iitm.ac.in}
\affiliation{Department of Physics, Indian Institute of Technology Madras, Chennai 600036, India}
\affiliation{Center for Quantum Information, Communication and Computation (CQuICC), Indian Institute of Technology Madras, Chennai 600036, India}

\begin{abstract}
We investigate the cluster Ising chain with an additional nearest-neighbor interaction using tensor-network methods and a weak-coupling field theory. Without the interaction, the Jordan-Wigner transformation decomposes the model into a triplet of identical Majorana chains related by an exact $O(3)$ flavor symmetry. Their common mass vanishes at the $SU(2)_2$ Wess-Zumino-Novikov-Witten critical point with central charge $c = 3/2$ separating the symmetry-protected topological cluster phase from a ferromagnet. The interaction reduces $O(3)$ to its cyclic subgroup $C_3$, splitting the triplet into a singlet and a doublet whose gaps close separately, producing a \textit{critical bifurcation} into Ising ($c = 1/2$) and Gaussian ($c = 1$) critical lines. A second ferromagnetic phase opens between these critical lines for repulsive interactions and a disordered phase for attractive ones. The Gaussian line then separates two Landau-incompatible ferromagnets, realizing a \textit{deconfined quantum critical} line with emergent $O(2)$ symmetry, along which our independently extracted exponents vary continuously yet satisfy the parameter-free relation $\beta = (2\nu - 1)/4$ of the eight-vertex weak universality class. At stronger repulsion this line opens into an extended gapless floating phase with incommensurate algebraic correlations, entered through Berezinskii-Kosterlitz-Thouless transitions. All of these phases and the transitions between them are captured by the weak-coupling theory. Beyond its regime of validity, our simulations reveal a translation-symmetry-breaking antiferromagnet reached through first-order transitions.
\end{abstract}

\date{\today}

\maketitle

\section{Introduction}
\label{sec:intro}

The Landau-Ginzburg paradigm has long provided the organizing principle for phase transitions~\cite{landau_1937, ginzburg_landau_1950}. Phases are classified by the symmetries they break and characterized by local order parameters, whose long-wavelength fluctuations become critical at a continuous transition. Combined with the renormalization group (RG)~\cite{wilson_kogut_1974, cardy_1996, fisher_rmp_1998}, this framework organizes critical behavior into universality classes largely independent of microscopic details~\cite{sachdev_2011}. Powerful as it is, it does not encompass all possible transitions. 
Topologically ordered states and symmetry-protected topological (SPT) phases carry no local order parameter, so that transitions involving them cannot be described by spontaneous breaking of a symmetry alone~\cite{wegner_jmp_1971, thouless_prl_1982, haldane_prl_1983, wen_book_2007, wen_rmp_2017}.
Berezinskii-Kosterlitz-Thouless (BKT) transitions are instead driven by the unbinding of topological defects and exhibit an exponentially diverging correlation length~\cite{berezinsky_jetp_1970, kosterlitz_jpc_1973, kosterlitz_jpc_1974, jose_prb_1977}. Another departure from the Landau-Ginzburg paradigm, central to this work, is \textit{deconfined quantum criticality} (DQC), in which two conventional symmetry-breaking phases are connected by a continuous transition that cannot be described by fluctuations of either local order parameter~\cite{senthil_science_2004, senthil_prb_2004, wang_prx_2017, senthil_review_2024}.

\begin{figure}[htb]
\centering
\begin{overpic}[width=\linewidth]{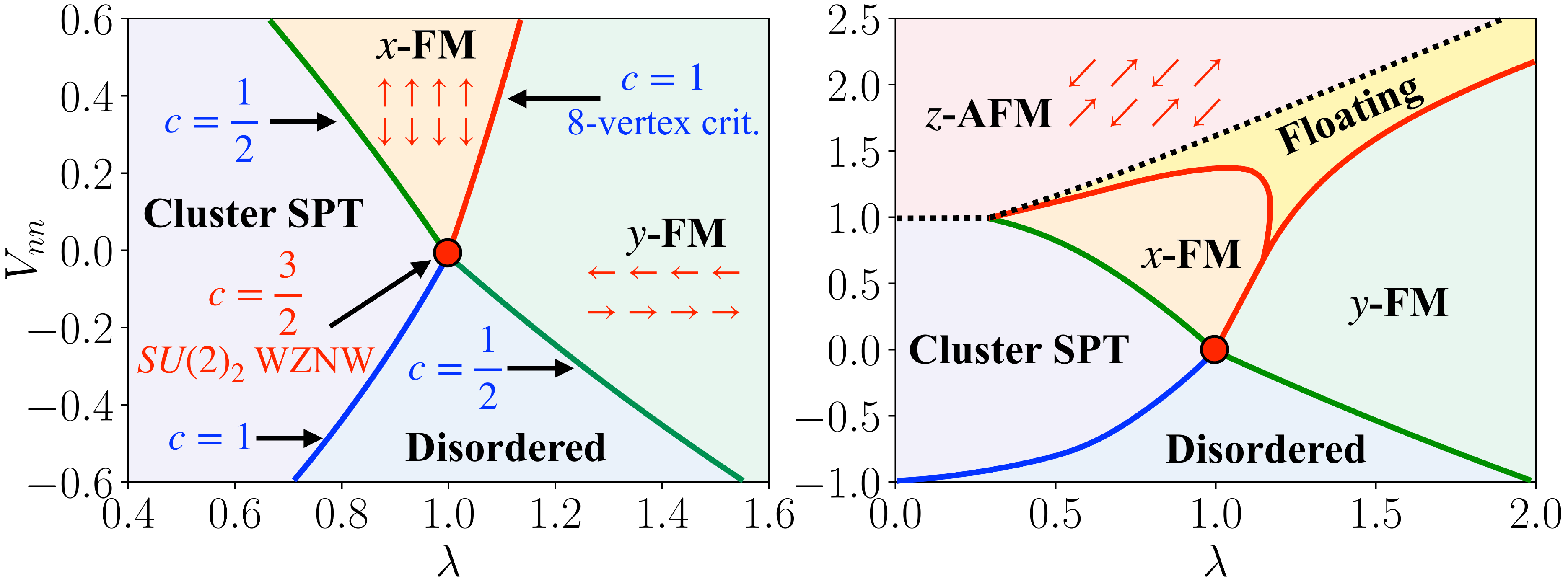}
\put(8.5, 33){\footnotesize (a)} 
\put(57.5, 33){\footnotesize (b)} 
\end{overpic}
\caption{(Color online.) Schematic illustration of the critical bifurcation in the interacting cluster Ising model and the phases it produces, in the plane of the Ising coupling $\lambda$ and the interaction strength $V_{nn}$ (Eqs.~\eqref{eq:H0} and~\eqref{eq:H}). (a) Near the non-interacting critical point (red dot), a cluster SPT phase and a ferromagnet ($y$-FM) meet at the $SU(2)_2$ WZNW point with central charge $c = 3/2$. Interactions split this point into an Ising line with $c = 1/2$ (green) and a Gaussian line with $c = 1$ (blue and red), which bound an intervening phase: a second ferromagnet ($x$-FM) for repulsive interactions, breaking a symmetry incompatible with that of the $y$-FM, and a disordered phase for attractive ones. The red line separating the two ferromagnets is the DQC line, belonging to the eight-vertex weak universality class. (b) The same over a wider range of couplings. At stronger repulsion the DQC line widens into a gapless floating phase with incommensurate correlations, which separates the ferromagnets from a translation-symmetry-breaking antiferromagnet ($z$-AFM) entered through first-order transitions (dotted). The boundaries are drawn schematically from the numerical calculations performed in this work.}
\label{fig:schem}
\end{figure}

The idea of DQC originated with the transition between the N\'eel antiferromagnet and the valence-bond solid in two dimensions (2D). These phases break unrelated symmetries, and neither can be reached from the other by further symmetry breaking. Within Landau-Ginzburg theory, such \textit{Landau-incompatible} phases can be separated only by a first-order transition, a coexistence region, or an intervening phase.  It was theorized, however, that a direct continuous transition is possible, driven by fractionalized spinons coupled to an emergent gauge field, which are confined on both sides and deconfined only at the critical point~\cite{senthil_science_2004, senthil_prb_2004}.
Numerical studies of lattice models initially provided extensive support for this scenario~\cite{sandvik_prl_2007, melko_prl_2008, kaul_prl_2012, sato_prl_2017, kaul_arcmp_2013}. However, many of these results have since been questioned, as larger-scale simulations reveal persistent scaling violations and exponents that drift with system size, and the transitions are now widely interpreted as weakly first order or pseudocritical~\cite{kuklov_prl_2008, nahum_prx_2015, zhao_np_2019, nahum_prb_2020, chester_prl_2024}. The nature of these transitions in 2D therefore remains debated.

One dimension (1D) offers a more controlled setting in this respect. Continuous transitions in 1D quantum systems, or equivalently in 2D classical statistical-mechanical models, are commonly described by conformal field theories (CFTs), whose central charges and operator scaling dimensions sharply constrain the universality class~\cite{difrancesco_1997}. Analytically, continuum-limit analysis, such as bosonization,  can connect a microscopic model to its low-energy field theory and predict its critical properties~\cite{gogolin_2004, giamarchi_2003, mussardo_book_2020}. Numerical methods based on matrix-product states (MPS)~\cite{schollwock_aop_2011, orus_aop_2014}, in turn, provide controlled access to both gapped and critical systems, allowing the theoretical predictions to be tested directly with high numerical accuracy. Continuous transitions between Landau-incompatible phases have consequently been established in several 1D quantum models and 2D classical systems~\cite{jiang_prb_2019, roberts_prb_2019, mudry_prb_2019, zhang_prl_2023, prakash_prl_2025, vishnu_prl_2025, pronk_prb_2025}.

A particularly well-controlled platform for exploring these questions is the cluster Ising model (CIM), a quantum spin chain in which a three-site cluster interaction competes with a nearest-neighbor Ising coupling~\cite{son_qip_2011, smacchia_pra_2011, son_epl_2011}. Under the Jordan-Wigner transformation, the model becomes a chain of non-interacting fermions and is exactly solvable. When the cluster interaction dominates, the ground state is the 1D cluster state familiar from measurement-based quantum computation~\cite{briegel_prl_2001}, an SPT phase diagnosed by non-local string order. When the Ising interaction dominates, the system develops conventional ferromagnetic (FM) order. The continuous transition between these phases already lies beyond a description based solely on a local symmetry-breaking order parameter. Furthermore, this criticality is richer than a symmetry analysis alone would suggest~\cite{smacchia_pra_2011, lahtinen_prl_2015}. Although only a single Ising symmetry is broken across it, the critical point carries central charge $c = 3/2$ and realizes the $SU(2)_2$ Wess-Zumino-Novikov-Witten (WZNW) theory~\cite{wess_plb_1971, novikov_rms_1982, witten_cmp_1984, knizhnik_npb_1984, difrancesco_1997, gogolin_2004} rather than the ordinary Ising universality class. Beyond its theoretical appeal, the model is a natural target for programmable quantum simulators. Different variants of cluster Hamiltonians have already been implemented on superconducting processors with tens to a hundred qubits, together with the string-order, edge, and entanglement properties needed to characterize them~\cite{zhang_nature_2022, jin_nature_2025, shen_npjqi_2025, tan_commphys_2026}.

In this work, we show that the CIM, away from its free-fermion limit, hosts a DQC line. We first identify the hidden symmetry underlying the aforementioned $c = 3/2$ criticality present in the free-fermion limit. The Jordan-Wigner transformation decomposes the chain into three identical Majorana species, and the rotations among them form an $O(3)$ symmetry that is exact throughout the non-interacting model. We then add a nearest-neighbor interaction that preserves the microscopic symmetries of the spin model but breaks this hidden $O(3)$ symmetry, and spoils integrability by rendering the Jordan-Wigner fermions interacting. Combining tensor-network simulations with a weak-coupling field theory, we find that the interaction splits the $c = 3/2$ critical point into two distinct critical lines, one of Ising type with $c = 1/2$ and one Gaussian with $c = 1$, a \textit{critical bifurcation} from which everything else follows.

Microscopically, this bifurcation originates from the reduction of $O(3)$ to its cyclic subgroup $C_3$, under which the Majorana triplet decomposes into a singlet and a real doublet with independent masses. Closing the singlet gap produces the Ising line, and closing the doublet gap the Gaussian one. The two critical lines bound an intervening phase (Fig.~\ref{fig:schem}(a)), an FM phase distinct from the original one for repulsive interactions and a disordered phase for attractive ones. The Gaussian line then separates two FM phases whose broken symmetries are incompatible, so that Landau-Ginzburg theory forbids a continuous transition between them. We find one nonetheless. This is the DQC line announced above, at which an emergent $O(2)$ symmetry rotates the two order parameters into one another, and by extracting the critical exponents we show that it belongs to the eight-vertex weak universality class~\cite{baxter_aop_1972, baxter_aop_1972b, kadanoff_aop_1979, baxter_book_1985}. Driving the repulsion further, the DQC line widens into an extended gapless floating phase with incommensurate correlations~\cite{japaridze_jltp_1979, pokrovsky_prl_1979} (Fig.~\ref{fig:schem}(b)). The weak-coupling theory predicts each of these features and agrees with the numerical simulations. At still stronger repulsion, beyond the controlled weak-coupling regime, we find a translation-symmetry-breaking antiferromagnetic (AFM) phase separated from its neighboring phases by first-order transitions.

The paper is organized as follows. In Sec.~\ref{sec:background}, we introduce the model, discuss its symmetries, and describe the numerical methods. In Secs.~\ref{sec:weak_nn} and \ref{sec:dqcl}, we investigate the weak-interaction regime. We first establish the critical bifurcation and explain its microscopic origin in Sec.~\ref{sec:weak_nn}, and then analyze the DQC line separating the two Landau-incompatible FM phases in Sec.~\ref{sec:dqcl}. In Sec.~\ref{sec:strong_nn}, we turn to stronger interactions and examine the emergence of the floating and AFM phases. We summarize our results in Sec.~\ref{sec:conclu}. A complete, self-contained microscopic derivation of the field-theoretical description is provided in Appendix~\ref{app:weak_coupling_derivation}, while the main text retains only the results essential to the discussion.

\section{Background and setup}
\label{sec:background}

In this section, we introduce the CIM, review its phases and symmetries, and
show how the Jordan-Wigner transformation exposes a hidden $O(3)$ symmetry underlying its
$c = 3/2$ criticality. We then define the interacting model studied in this work and describe
the numerical methods used to analyze it.

\subsection{The cluster Ising model}
\label{subsec:cim}

We start with the 1D CIM, described by the Hamiltonian~\cite{son_qip_2011, smacchia_pra_2011, son_epl_2011}
\begin{align}
\hat{H}_0 = &- J \sum_j \hat{\sigma}^x_j \hat{\sigma}^z_{j + 1} \hat{\sigma}^x_{j + 2}
- \lambda \sum_j \hat{\sigma}^y_j \hat{\sigma}^y_{j + 1},
\label{eq:H0}
\end{align}
where $\hat{\sigma}^{\alpha}_j$, with $\alpha = x, y, z$, are the Pauli operators acting on site $j$. The first term corresponds to a three-spin cluster interaction involving sites $j$, $j + 1$, and $j + 2$. Throughout the paper, we set $J = 1$ to fix the unit of energy. The second term represents a nearest-neighbor Ising interaction along the $y$ direction with coupling strength $\lambda$. Under the Jordan-Wigner transformation, $\hat{H}_0$ can be mapped to a chain of non-interacting spinless fermions and solved exactly~\cite{smacchia_pra_2011}. 

The Hamiltonian~\eqref{eq:H0} is invariant under a $\mathbb{Z}_2 \times \mathbb{Z}_2$ symmetry, generated by the parity transformation
\begin{align}
    \mathcal{P}: \ \hat{\sigma}^{x, y} \rightarrow -\hat{\sigma}^{x, y}, \ \hat{\sigma}^z \rightarrow \hat{\sigma}^z
    \label{eq:P}
\end{align}
and the complex conjugation in the $\hat{\sigma}^z$ basis
\begin{align}
    \mathcal{K}: \ \hat{\sigma}^y \rightarrow -\hat{\sigma}^y, \ \hat{\sigma}^{x, z} \rightarrow \hat{\sigma}^{x, z}.
    \label{eq:K}
\end{align}
Beyond the manifest discrete symmetries of the spin model, as we discuss below, $\hat{H}_0$ also respects a hidden but exact continuous $O(3)$ symmetry~(cf.~\cite{lahtinen_prl_2015}).

In the $\lambda = 0$ limit, the system is in a cluster state stabilized by the set of stabilizers $\hat{K}_j = \hat{\sigma}^x_{j - 1} \hat{\sigma}^z_{j} \hat{\sigma}^x_{j + 1}$, such that $\hat{K}_j \ket{\psi} = +1 \ket{\psi}$ for all $j$~\cite{briegel_prl_2001, smacchia_pra_2011, son_qip_2011}. This is an example of a bosonic SPT state, 
and is, in fact, the fixed-point limit of the well-known spin-1 Affleck-Kennedy-Lieb-Tasaki (AKLT)~\cite{affleck_prl_1987, affleck_cmp_1988} state in disguise, related to it via a local unitary circuit~\cite{verresen_prb_2017}.
 For non-zero $\lambda$, this SPT phase, which we refer to as the \textit{cluster} phase, persists up to $\lambda = 1$. It is protected by the $\mathbb{Z}_2 \times \mathbb{Z}_2$ symmetry introduced above~\cite{son_qip_2011}.
For $\lambda > 1$, the system enters an FM phase in which the spins align along either the positive or the negative $y$-direction~\cite{son_epl_2011, smacchia_pra_2011}. In this phase, $\mathcal{P}$ and $\mathcal{K}$ are both broken while their product $\mathcal{PK}$ remains a good symmetry, corresponding to $\mathbb{Z}_2 \times \mathbb{Z}_2 \to \mathbb{Z}_2$ breaking as in an ordinary ferromagnet. We refer to this phase as the $y$-FM phase.

The two phases identified above are distinguished by two complementary order parameters. Since the cluster phase is an SPT phase, it carries no local order parameter and is diagnosed by the non-local string order parameter~\cite{smacchia_pra_2011, son_qip_2011}
\begin{align}
O_S = &\lim_{R \rightarrow \infty} \left| \Braket{\hat\sigma^x_{i} \hat\sigma^y_{i + 1} \left(\prod_{k = i + 2}^{i + R - 2} \hat\sigma^z_{k}\right) \hat\sigma^y_{i + R - 1}\hat\sigma^x_{i + R}} \right|,
\label{eq:string_order}
\end{align}
which takes a non-zero value in the cluster phase and vanishes identically in the $y$-FM phase. The $y$-FM phase, on the other hand, is captured by the conventional local order parameter, i.e.,  the magnetization along the $y$-direction
\begin{align}
m_y =  \frac{1}{L} \sum_j  \langle \hat\sigma^y_j \rangle,
\label{eq:my}
\end{align}
with $L$ being the system size,
which vanishes throughout the cluster phase and becomes non-zero for $\lambda > 1$~\cite{son_epl_2011}. The two order parameters are thus mutually exclusive, and both vanish continuously as $\lambda \to 1$ from either side, consistent with a continuous quantum phase transition (QPT) separating the two phases.

The QPT at $\lambda = 1$ is considerably richer than a conventional Ising transition. Although the transition from the cluster SPT phase to the $y$-FM phase involves the spontaneous breaking of a single $\mathbb{Z}_2$ symmetry, it is described by a CFT with central charge $c = 3/2$~\cite{smacchia_pra_2011}, three times the $c = 1/2$ of a single critical Ising chain. This tripling is not accidental. It reflects a continuous symmetry that is hidden at the level of the spin degrees of freedom, revealed only by a non-local transformation.

\subsection{Hidden $O(3)$ symmetry}
\label{subsec:hidden_o3}

The hidden symmetry becomes manifest after performing the Jordan-Wigner transformation~\cite{mbeng_scipostlect_2024},
\begin{align}
\hat{\sigma}^x_j &= (-1)^j \prod_{l < j} \left( 1 - 2 \hat{n}_l \right)
\left( \hat{c}^\dagger_j + \hat{c}_j \right), \notag\\
\hat{\sigma}^y_j &= i (-1)^j \prod_{l < j} \left( 1 - 2 \hat{n}_l \right)
\left( \hat{c}^\dagger_j - \hat{c}_j \right), \notag \\
\hat{\sigma}^z_j &= 1 - 2 \hat{n}_j,
\label{eq:jw}
\end{align}
where $\hat{n}_j = \hat{c}_j^\dagger\hat{c}_j$. Under this mapping, the Hamiltonian~\eqref{eq:H0} becomes
\begin{align}
\hat{H}_0 = &-  \sum_j \left( \hat{c}^\dagger_j - \hat{c}_j \right) \left( \hat{c}^\dagger_{j + 2} + \hat{c}_{j + 2} \right) \notag\\
&- \lambda \sum_j \left( \hat{c}^\dagger_j + \hat{c}_j \right) \left( \hat{c}^\dagger_{j + 1} - \hat{c}_{j + 1} \right).
\label{eq:H0_fermion}
\end{align}
Introducing the Majorana operators $\hat{\zeta}_j = \hat{c}_j^\dagger + \hat{c}_j$ and $\hat{\eta}_j = i(\hat{c}_j^\dagger - \hat{c}_j)$, the Hamiltonian can be written compactly as
\begin{align}
\hat{H}_0 = i\sum_j \left[\hat{\eta}_j\hat{\zeta}_{j + 2} + \lambda\,\hat{\zeta}_j\hat{\eta}_{j + 1} \right].
\label{eq:H0_majorana}
\end{align}
Equation~\eqref{eq:H0_majorana} reveals a coupling pattern that closes into three independent Majorana chains. Defining
\begin{align}
 \hat{\zeta}^a_k = \hat{\zeta}_{3k + a - 1}, \quad \hat{\eta}^a_k = \hat{\eta}_{3k + a},
 \label{eq:flavor_define}
\end{align}
for $a = 1, 2, 3$, the Hamiltonian assumes the form
\begin{align}
\hat{H}_0 = i\sum_{a = 1}^{3}\sum_k \left[\hat{\eta}^a_k\hat{\zeta}^a_{k + 1} + \lambda\,\hat{\zeta}^a_k\hat{\eta}^a_k \right].
\label{eq:H0_flavors}
\end{align}
The continuum limit of Eq.~\eqref{eq:H0_flavors} is the three-flavor Majorana field theory~\cite{gogolin_2004},
\begin{align}
\hat{H}_0 \to i \sum_{a = 1}^{3}\int dx\left(v_0\,\eta^a\partial_x\zeta^a - m_0\,\eta^a\zeta^a\right),
\label{eq:continuum_free}
\end{align}
where $v_0$ and $m_0$ are the common velocity and mass. The explicit continuum normalization in Appendix~\ref{app:free_continuum} gives $v_0 = 6a_0$ and $m_0 = 2(\lambda - 1)$, with $a_0$ being the lattice spacing. The three Majorana flavors can be rotated into one another, revealing the exact $O(3)$ symmetry of the CIM.
Consequently, all three masses vanish together at $\lambda = 1$, where the three massless Majoranas realize the $SU(2)_2 \simeq SO(3)_1$ WZNW CFT with $c = 3/2$~\cite{wess_plb_1971, novikov_rms_1982, witten_cmp_1984, knizhnik_npb_1984, difrancesco_1997, gogolin_2004}. 
This $O(3)$ symmetry is already manifest in the lattice Majorana Hamiltonian,
Eq.~\eqref{eq:H0_flavors}, and holds for all $\lambda$. It is therefore exact rather than
emergent. The interaction introduced below spoils this symmetry explicitly.

\subsection{Interacting cluster Ising model}
\label{subsec:interacting_model}

The central object of this work is the interacting cluster Ising model, obtained by
supplementing Eq.~\eqref{eq:H0} with a nearest-neighbor interaction along the $z$ direction,
\begin{align}
\hat{H} = \hat{H}_0 + V_{nn} \sum_j \hat{\sigma}^z_j \hat{\sigma}^z_{j + 1},
\label{eq:H}
\end{align}
where $V_{nn}$ denotes the interaction strength. The $\hat{\sigma}^z_j \hat{\sigma}^z_{j + 1}$ term respects $\mathcal{P}$ and
$\mathcal{K}$, but not the hidden $O(3)$ symmetry. Using $\hat{\sigma}^z_j =
-i\hat{\zeta}_j\hat{\eta}_j$ and the flavor assignment of Eq.~\eqref{eq:flavor_define}, the
three bonds within each unit cell give
\begin{align}
V_{nn} \sum_j \hat{\sigma}^z_j & \hat{\sigma}^z_{j + 1} = -V_{nn}\sum_k\big[
(\hat{\zeta}^2_k\hat{\eta}^2_k)
(\hat{\eta}^1_k\hat{\zeta}^3_k) \notag \\
+\, & (\hat{\zeta}^3_k\hat{\eta}^3_k)
(\hat{\eta}^2_k\hat{\zeta}^1_{k + 1})
+ (\hat{\zeta}^1_{k + 1}\hat{\eta}^1_{k + 1})
(\hat{\eta}^3_k\hat{\zeta}^2_{k + 1})\big],
\label{eq:zz_interaction}
\end{align}
so that every term is quartic in the Majorana operators, couples all three flavors, and 
the model is therefore no longer exactly solvable.
What replaces the $c = 3/2$ critical point at non-zero $V_{nn}$ is the
problem addressed in the rest of this paper.
We approach the problem both analytically and
numerically. Analytically, we develop a weak-coupling field theory around the non-interacting
critical point $(\lambda, V_{nn}) = (1, 0)$ using Majorana fields and bosonization, derived in
detail in Appendix~\ref{app:weak_coupling_derivation}. Numerically, we use the tensor-network methods described below.

We obtain the ground states of the Hamiltonian~\eqref{eq:H} using MPS
methods~\cite{schollwock_aop_2011,orus_aop_2014}, specifically the variational uniform MPS (VUMPS) algorithm~\cite{zaunerstauber_prb_2018,vanderstraeten_scipost_2019}, a
tangent-space approach~\cite{haegeman_prl_2011} that directly optimizes a uniform MPS (uMPS) in
the thermodynamic limit. 

To map out the phase diagram and locate the transitions, we examine the bipartite entanglement
entropy of the optimized uMPS,
\begin{align}
\mathcal{S} = -\sum_{\alpha} \Lambda_{\alpha}^2 \log \Lambda_{\alpha}^2,
\label{eq:ee}
\end{align}
where $\Lambda_{\alpha}$ are the Schmidt coefficients across a given bond. At a critical point,
a uMPS with finite bond dimension $\chi$ is limited to a finite, $\chi$-dependent correlation
length $\xi_{\chi}$~\cite{tagliacozzo_prb_2008,pollmann_prl_2009}, from which the central charge
can be extracted via the finite-entanglement scaling
relation~\cite{callan_geometric_1994, vidal_PRL_2003, calabrese_entanglement_2004}
\begin{align}
\mathcal{S}(\chi) = \frac{c}{6}\log \xi_{\chi} + b',
\label{eq:calabrese_cardy}
\end{align}
where $b'$ is a non-universal constant and $\xi_\chi = -1/\log|\epsilon_1/\epsilon_0|$, with
$\epsilon_0$ and $\epsilon_1$ the dominant and subdominant eigenvalues of the uMPS transfer
matrix, respectively. 

Furthermore, to complement finite-entanglement scaling analysis of critical transitions using VUMPS, 
we also use, whenever needed, finite-size density-matrix renormalization group (DMRG)~\cite{white_prl_1992, white_prb_1993, schollwock_aop_2011,orus_aop_2014}, and perform standard finite-size scaling analysis across critical phase transitions to determine the critical exponents $\beta$ and $\nu$.

\section{Critical bifurcation under weak interactions}
\label{sec:weak_nn}

\begin{figure}[t]
    \centering
    \includegraphics[width=\linewidth]{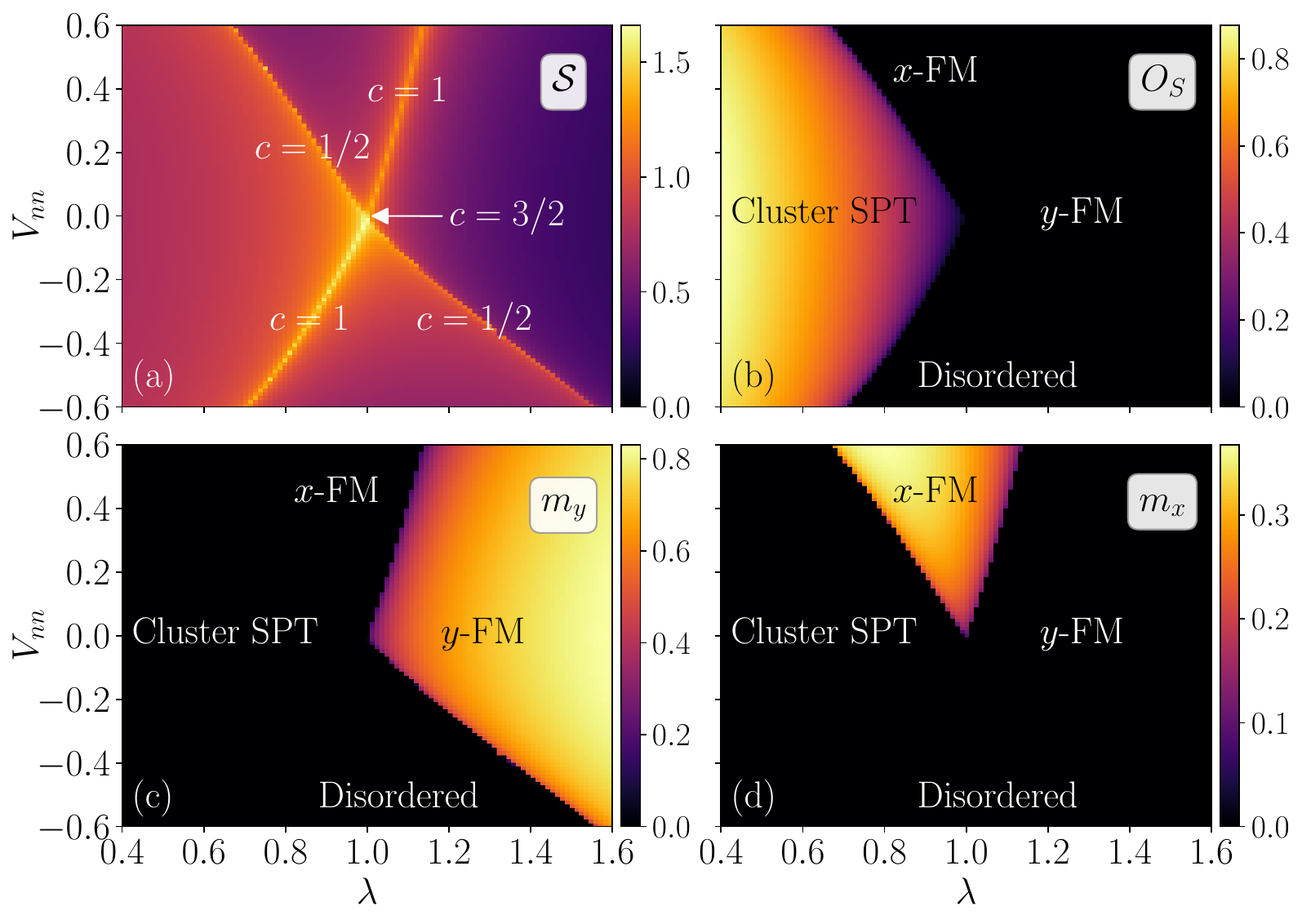}
    \caption{(Color online.) Phase diagram of the interacting CIM described by Eq.~\eqref{eq:H} in the vicinity of the non-interacting critical point $(\lambda, V_{nn}) = (1, 0)$. We show the bipartite entanglement entropy $\mathcal{S}$ (a), the string order parameter $O_S$ (b), and the magnetizations $m_y$ (c) and $m_x$ (d), obtained using VUMPS with uMPS bond dimension $\chi = 128$. For $V_{nn} > 0$, an $x$-FM phase with non-zero $m_x$ emerges between the cluster and $y$-FM phases; for $V_{nn} < 0$, these phases are instead separated by a disordered phase, and thus the non-interacting critical point at $(\lambda, V_{nn}) = (1, 0)$ bifurcates into two transition lines on either side of the $V_{nn} = 0$ axis. The central charges labeled in (a) are confirmed via the finite-entanglement scaling analysis in Fig.~\ref{fig:cuts_ent_nn}.}
    \label{fig:phase_nn}
\end{figure}

We first treat $V_{nn}$ as a weak perturbation to the non-interacting CIM and focus on the vicinity of the $(\lambda, V_{nn}) = (1, 0)$ critical point. Figure~\ref{fig:phase_nn} shows the bipartite entanglement entropy and the relevant order parameters in the $(\lambda, V_{nn})$-plane. In addition to the string order parameter $O_S$ and the $y$-magnetization $m_y$, we consider the magnetization along the $x$-direction,
\begin{align}
m_x =  \frac{1}{L} \sum_j \Braket{\hat{\sigma}_j^x}.
\label{eq:mx}
\end{align}
At $V_{nn} = 0$, the cluster and $y$-FM phases meet at the $c = 3/2$ critical point at $\lambda = 1$. For any non-zero $V_{nn}$, this single critical point bifurcates into two critical lines on either side of the $V_{nn} = 0$ axis. The direct cluster $\leftrightarrow$ $y$-FM transition is therefore replaced by two successive transitions separated by an intervening phase, whose nature depends on the sign of $V_{nn}$.

The different phases can be identified from the order parameters shown in Figs.~\ref{fig:phase_nn}(b)-(d). For $V_{nn} > 0$, the phase intervening between the cluster and $y$-FM phases is characterized by $m_x \neq 0$, while both $O_S$ and $m_y$ vanish, indicating ferromagnetic order along the $x$-direction. We therefore refer to it as the $x$-FM phase. For $V_{nn} < 0$, the intervening phase instead has vanishing $O_S$, $m_x$, and $m_y$, together with a finite polarization along the positive $z$-direction (not shown). Since this phase preserves the microscopic $\mathbb{Z}_2 \times \mathbb{Z}_2$ symmetry and exhibits no spontaneous symmetry breaking, it is disordered in the Landau sense. The two signs of $V_{nn}$ therefore split the original $c = 3/2$ transition in qualitatively distinct ways.

\begin{figure}[t]
    \centering
    \includegraphics[width=\linewidth]{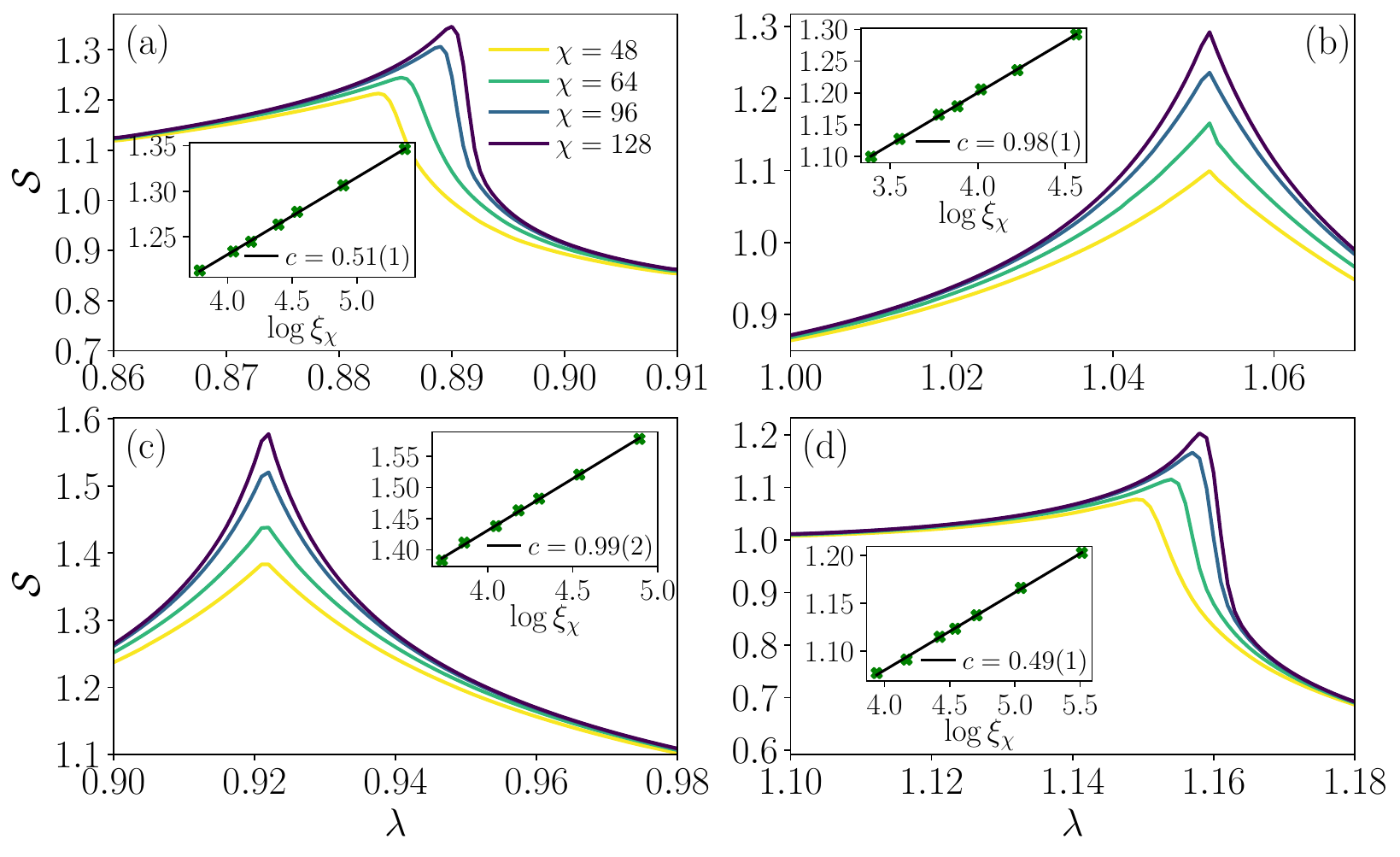}
    \caption{(Color online.) Bipartite entanglement entropy across the four bifurcated transition lines of the interacting CIM for different bond dimensions $\chi \in [48, 128]$. The upper panels correspond to $V_{nn} = 0.2$ and show the cluster $\leftrightarrow$ $x$-FM (a) and $x$-FM $\leftrightarrow$ $y$-FM (b) transitions, whereas the lower panels correspond to $V_{nn} = -0.2$ and show the cluster $\leftrightarrow$ disordered (c) and disordered $\leftrightarrow$ $y$-FM (d) transitions. The entropy maxima locate the respective transition points for each bond dimension. The insets show the finite-entanglement scaling of these maxima with the corresponding correlation lengths $\xi_{\chi}$ according to Eq.~\eqref{eq:calabrese_cardy}. The extracted central charges are consistent with $c = 1/2$ Ising criticality for the transitions in panels (a) and (d), and with $c = 1$ criticality for those in panels (b) and (c).}
    \label{fig:cuts_ent_nn}
\end{figure}

Having identified the phases surrounding the bifurcation, we next characterize the four transition lines emanating from the $(\lambda, V_{nn}) = (1, 0)$ critical point. The behavior of the bipartite entanglement entropy and the corresponding order parameters indicates that all four transitions are continuous. To determine their nature, we perform finite-entanglement scaling at a representative cut through the phase diagram. Figure~\ref{fig:cuts_ent_nn} shows the bipartite entanglement entropy as a function of $\lambda$ for several bond dimensions, with $V_{nn} = 0.2$ and $V_{nn} = -0.2$. Following Refs.~\cite{kjall_prb_2013, tsitsishvili_prb_2022}, we locate, for each transition, the $\chi$-dependent maxima of the bipartite entanglement entropy, evaluate the corresponding correlation length $\xi_{\chi}$ at each maximum, and extract the central charge by fitting the resulting $\mathcal{S}(\chi)$ and $\xi_{\chi}$ to Eq.~\eqref{eq:calabrese_cardy}.

The resulting scaling analyses are shown in the insets of Fig.~\ref{fig:cuts_ent_nn}. For $V_{nn} > 0$, the transition between the cluster and $x$-FM phases is governed by Ising criticality with $c = 1/2$~\footnote{This is an Ising transition between the cluster SPT phase and an Ising ferromagnet, at which the protecting $\mathbb{Z}_2\times\mathbb{Z}_2$ symmetry remains unbroken, as in Ref.~\cite{verresen_prx_2021}. It may, therefore, be of the symmetry-enriched kind~\cite{verresen_prx_2021, mondal_prb_2023}, a question we leave for future work.}, 
whereas the transition between the $x$-FM and $y$-FM phases is described by a $c = 1$ critical theory. For $V_{nn} < 0$, the cluster $\leftrightarrow$ disordered transition is characterized by $c = 1$ criticality, while the disordered $\leftrightarrow$ $y$-FM transition belongs to the Ising universality class with $c = 1/2$. Thus, the nearest-neighbor interaction splits the original $c = 3/2$ critical point into $c = 1$ and $c = 1/2$ critical lines, whose ordering is reversed upon changing the sign of $V_{nn}$. Below, we show that this bifurcation arises because the interaction reduces the $O(3)$ flavor symmetry of the parent critical point at $(\lambda, V_{nn}) = (1, 0)$ to a residual cyclic symmetry $C_3$, under which the Majorana triplet decomposes into a singlet and a doublet.

Even in the presence of the $\hat{\sigma}^z_j \hat{\sigma}^z_{j + 1}$ interaction (Eq.~\eqref{eq:zz_interaction}), the Hamiltonian remains invariant under the cyclic transformation
\begin{align}
\hat{\zeta}^1_k
\rightarrow \hat{\zeta}^2_k
\rightarrow \hat{\zeta}^3_k
\rightarrow \hat{\zeta}^1_{k + 1},
\end{align}
and likewise for $\hat{\eta}^a_k$. The cell shift in the final step generates only derivative corrections in the continuum and therefore does not affect the leading zero-momentum mass terms. At this level, one-site translation acts as a cyclic permutation of the three flavors, represented by
\begin{align}
C =
\begin{pmatrix}
0 & 1 & 0\\
0 & 0 & 1\\
1 & 0 & 0
\end{pmatrix},
\qquad
C^3 = \mathbb{I}_3.
\label{eq:cyclic_flavor_matrix}
\end{align}

This residual $C_3$ symmetry constrains how the interaction can lift the degeneracy of the three Majorana flavors. Combining the common mass of the non-interacting theory (Eq.~\eqref{eq:continuum_free}) with the quadratic corrections generated by $V_{nn}$, we write the renormalized zero-momentum mass term as
\begin{align}
H_M = -i\int dx\,
\boldsymbol{\eta}^T M\boldsymbol{\zeta},
\label{eq:continuum_mass_matrix}
\end{align}
where $\boldsymbol{\eta} = (\eta^1, \eta^2, \eta^3)^T$ and $\boldsymbol{\zeta} = (\zeta^1, \zeta^2, \zeta^3)^T$. Hermiticity requires $M$ to be real, while invariance under the cyclic transformation requires $C^TMC = M$. The most general matrix satisfying this constraint is (see Appendix~\ref{app:singlet_doublet})
\begin{align}
M = \widetilde{m}_0 \mathbb{I}_3
+ m_1(C + C^2) + m_2(C - C^2),
\label{eq:mass_matrix}
\end{align}
where $\widetilde{m}_0$ is the interaction-renormalized flavor-independent mass, while $m_1$ and $m_2$ are the coefficients of the symmetric and antisymmetric flavor-mixing terms generated by the interaction.

The $C_3$ symmetry organizes the three Majorana flavors into a singlet and a two-dimensional real doublet. The singlet corresponds to the symmetric combinations
$\eta_s = (\eta^1 + \eta^2 + \eta^3)/\sqrt{3}$ and
$\zeta_s = (\zeta^1 + \zeta^2 + \zeta^3)/\sqrt{3}$, while the combinations orthogonal to them span the doublet. Block diagonalization of the renormalized mass matrix in Eq.~\eqref{eq:mass_matrix}, detailed in Appendix~\ref{app:singlet_doublet}, then yields the gap-opening masses
\begin{align}
m_s = \widetilde{m}_0 + 2m_1,
\qquad
m_d = \widetilde{m}_0 - m_1,
\end{align}
for the singlet and doublet sectors, respectively.

Organizing different flavors of Majorana fermions into multiplets of an exact symmetry in this way is
familiar from the two-leg Heisenberg ladder. Two decoupled spin-$1/2$ Heisenberg chains realize
an $SU(2)_1 \times SU(2)_1$ theory, and coupling them splits the four Majorana fields into a triplet and a
singlet under $O(3) \otimes \mathbb{Z}_2$, where $O(3)$ rotates the triplet components into each other and the 
$\mathbb{Z}_2$ acts on the singlet~\cite{shelton_prb_1996, gogolin_2004}. 
The mechanism is the same here,
the cyclic $C_3$ symmetry yielding a doublet in place of the triplet. In both cases the
multiplet structure forbids mixing between the sectors, so that they can become critical separately.

In the present case, the antisymmetric coefficient $m_2$ of Eq.~\eqref{eq:mass_matrix} mixes the two real components of the doublet without generating an additional gap. To leading order in the weak-coupling expansion, diagonalization of the quadratic Hamiltonian gives (see Appendix~\ref{app:dispersion_derivation})
\begin{align}
E_s^2(k) &= m_s^2 + v_s^2k^2, \notag\\
E_{d, \pm}^2(k) &= m_d^2 +
\left(v_dk \pm \sqrt{3}\,m_2\right)^2,
\label{eq:singlet_doublet_dispersions}
\end{align}
where $v_s$ and $v_d$ denote the renormalized velocities of the singlet and doublet sectors. The two doublet branches therefore have minima at
\begin{align}
q_\pm = \mp \frac{\sqrt{3}\,m_2}{v_d},
\label{eq:doublet_momentum_shift}
\end{align}
but share the same excitation gap $\Delta_d = |m_d|$. Evidently, $m_s = 0$ produces a single critical Majorana mode with $c = 1/2$, whereas $m_d = 0$ produces a critical Majorana doublet, equivalently a single Dirac mode, with $c = 1$.
The displacement of the minima away from zero momentum further implies that correlations in the
doublet sector oscillate at an incommensurate wavevector, both in the gapped phases on either
side and on the $c = 1$ line itself.

To locate the two critical lines, we expand about the non-interacting point $(\lambda, V_{nn}) = (1, 0)$. Normal ordering the lattice interaction and resolving its quadratic contribution into the singlet and doublet sectors, as detailed in Appendices~\ref{app:normal_ordering} and \ref{app:singlet_doublet}, gives
\begin{align}
m_s &= 2(\lambda - 1) + \frac{4V_{nn}}{\pi}
+ \mathcal{O}(V_{nn}^2), \notag\\
m_d &= 2(\lambda - 1) - \frac{2V_{nn}}{\pi}
+ \mathcal{O}(V_{nn}^2), \notag\\
m_2 &= -\frac{2V_{nn}}{\pi}
+ \mathcal{O}(V_{nn}^2).
\label{eq:split_masses_linear}
\end{align}
The conditions $m_s = 0$ and $m_d = 0$ therefore predict the critical lines
\begin{align}
\lambda_s &= 1 - \frac{2V_{nn}}{\pi}
+ \mathcal{O}(V_{nn}^2),&
\lambda_d &= 1 + \frac{V_{nn}}{\pi}
+ \mathcal{O}(V_{nn}^2).
\end{align}
They coincide at the $c = 3/2$ point for $V_{nn} = 0$ and exchange their order across it, reproducing the sequence of $c = 1/2$ and $c = 1$ transitions found numerically. For $V_{nn} = 0.2$, the leading-order expressions give $(\lambda_s, \lambda_d) \simeq (0.873, 1.064)$, compared with the numerical estimates $\sim~(0.89, 1.052)$ in Figs.~\ref{fig:cuts_ent_nn}(a) and \ref{fig:cuts_ent_nn}(b). For $V_{nn} = -0.2$, the leading-order estimate give $(\lambda_d, \lambda_s) \simeq (0.936, 1.127)$, compared with $\sim~(0.922, 1.16)$ in Figs.~\ref{fig:cuts_ent_nn}(c) and \ref{fig:cuts_ent_nn}(d). 
The agreement is already good at this leading order, with the deviations consistent with the neglected $\mathcal{O}(V_{nn}^2)$ corrections.

As a check on the above mechanism, replacing the nearest-neighbor coupling by a next-nearest-neighbor $\hat{\sigma}^z_j \hat{\sigma}^z_{j + 2}$ interaction reverses the singlet-doublet splitting and mirrors the two critical lines about the non-interacting axis, which we confirm in Appendix~\ref{subsec:vnnn_inversion}.

\section{Deconfined quantum critical line}
\label{sec:dqcl}

The most striking consequence of the critical bifurcation is the $c = 1$ critical line that directly separates the $x$-FM and $y$-FM phases for $V_{nn} > 0$. These are \textit{Landau-incompatible} symmetry-broken phases, characterized by distinct local order parameters: $m_x \neq 0$ and $m_y = 0$ in the $x$-FM phase, whereas $m_x = 0$ and $m_y \neq 0$ in the $y$-FM phase. 
The magnetizations inherit their transformation properties from the corresponding microscopic spin operators (Eqs.~\eqref{eq:P} and \eqref{eq:K}), giving
\begin{align}
\mathcal{P}:&\quad (m_x, m_y) \rightarrow (-m_x, -m_y), \notag\\
\mathcal{K}:&\quad (m_x, m_y) \rightarrow (m_x, -m_y).
\end{align}
Therefore, since $m_x$ is odd under $\mathcal{P}$ but even under $\mathcal{K}$, the $x$-FM phase preserves the residual symmetry $\{1, \mathcal{K}\}$. By contrast, $m_y$ is odd under both $\mathcal{P}$ and $\mathcal{K}$ but even under their product, so the $y$-FM phase preserves $\{1, \mathcal{PK}\}$. Neither residual symmetry group is a subgroup of the other, so neither phase can be obtained from the other by further symmetry breaking. Within conventional Landau-Ginzburg theory, which describes a continuous transition through fluctuations of a local order parameter, a direct continuous transition between these phases is therefore forbidden~\cite{senthil_science_2004, senthil_prb_2004, senthil_review_2024, wang_prx_2017}. The continuous $c = 1$ boundary found here instead realizes a one-dimensional DQC line~\cite{jiang_prb_2019, roberts_prb_2019, zhang_prl_2023, prakash_prl_2025, vishnu_prl_2025}.

\begin{figure}[htb]
    \centering
    \includegraphics[width=0.7\linewidth]{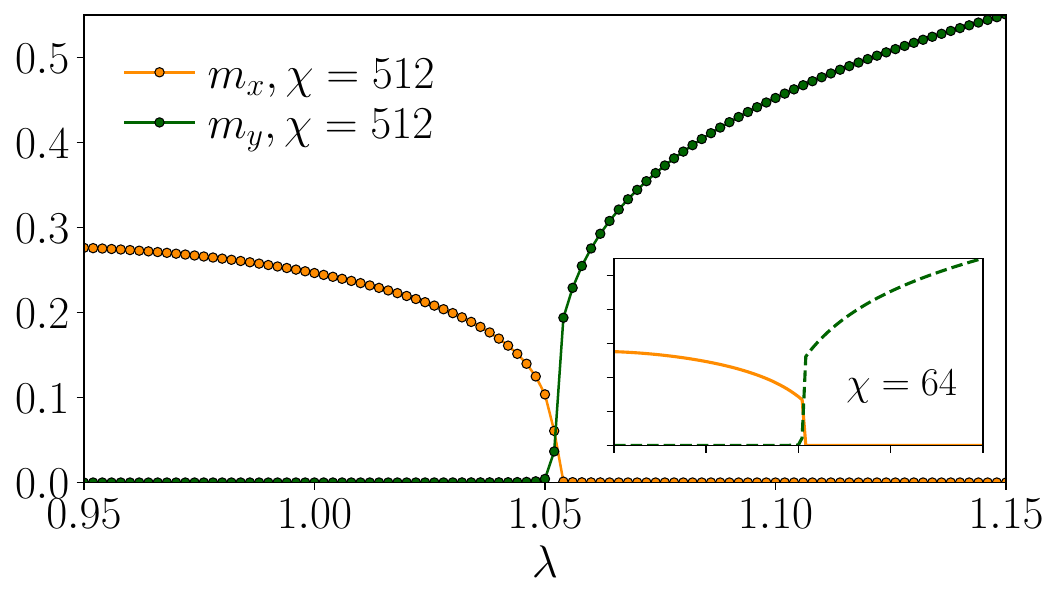}
    \caption{(Color online.) Magnetizations $m_x$ and $m_y$ as functions of $\lambda$ across the direct $c = 1$ transition between the $x$-FM and $y$-FM phases at fixed $V_{nn} = 0.2$, obtained using VUMPS with bond dimension $\chi = 512$. The inset shows the corresponding results for $\chi = 64$, where finite-entanglement effects make the transition appear considerably sharper.}
    \label{fig:mx_my_nn}
\end{figure}

Figure~\ref{fig:mx_my_nn} shows the simultaneous and continuous vanishing of $m_x$ and $m_y$ at the same critical point. The comparison between $\chi = 64$ and $\chi = 512$ shows that the continuous behavior is resolved only once the finite-entanglement correlation length is large enough. Together with the central charge $c = 1$ obtained in Sec.~\ref{sec:weak_nn}, these results establish a direct continuous transition with diverging correlation length between the two Landau-incompatible FM phases.

The simultaneous vanishing of $m_x$ and $m_y$ suggests that the two Landau-incompatible orders are related by an emergent continuous rotation at criticality. Phenomenologically, this can be described by a compact angular field $\vartheta(x) \equiv \vartheta(x) + 2\pi$, with
\begin{align}
m_x \sim \langle\cos\vartheta\rangle, \quad m_y \sim \langle\sin\vartheta\rangle.
\end{align}
The microscopic symmetries act as
\begin{align}
\mathcal{P}: \vartheta \rightarrow \vartheta + \pi, \ \ \mathcal{K}: \vartheta \rightarrow -\vartheta, \ \ \mathcal{PK}: \vartheta \rightarrow \pi - \vartheta,
\end{align}
reproducing the transformation properties of both magnetizations. The $x$-FM phase is obtained by locking $\vartheta$ at $0$ or $\pi$, whereas the $y$-FM phase corresponds to $\vartheta = \pi/2$ or $3\pi/2$. The leading anisotropy that selects between these two sets of configurations changes sign and vanishes at the transition. If the remaining symmetry-allowed anisotropies are irrelevant, the long-wavelength theory becomes invariant under $\vartheta \rightarrow \vartheta + \alpha$, yielding an emergent $U(1)$ symmetry that continuously rotates the two orders into one another. Together with the reflection $\mathcal{K}$, this enlarges the emergent symmetry to $O(2)$.

Hallmarks of a deconfined transition are already visible at this level~\cite{senthil_review_2024}.
The first is the emergent rotational symmetry, larger than that of either adjacent phase. The
second concerns the defects that drive the transition. The two $x$-FM ground states,
$\vartheta = 0$ and $\vartheta = \pi$, are invariant under $\mathcal{K}$, but a \textit{smooth domain
wall} connecting them must pass through $\vartheta = \pi/2$ or $3\pi/2$. These configurations are
charged under $\mathcal{K}$, so the wall carries a core of $y$-FM order. The roles are exchanged
in the $y$-FM phase, where the corresponding walls trap $\mathcal{PK}$ charge and hence $x$-FM
order. The relevant excitations on either side are therefore these charged domain walls rather
than the magnetizations themselves, and on approaching criticality their proliferation destroys
the surrounding order and drives the transition.

A phenomenological description of the transition can be provided by the compact-boson Hamiltonian~\cite{jose_prb_1977, lecheminant_npb_2002, mudry_prb_2019, prakash_prl_2025}
\begin{align}
H_{\mathrm{eff}} ={}& \frac{v}{2\pi}\int dx\left[ K(\partial_x\vartheta)^2 +\frac{1}{4K}(\partial_x\varphi)^2 \right] \notag\\
&- \int dx\left[ g_2\cos(2\vartheta) + g_4\cos(4\vartheta) + \ldots \right] \notag\\
&- \int dx\left[g_{3\varphi}\cos(3\varphi) + \ldots \right] \notag\\
&+ \frac{h}{2\pi}\int dx\,\partial_x\varphi.
\label{eq:sine_gordon}
\end{align}
Here, $\varphi$ is dual to $\vartheta$, both fields are $2\pi$ periodic, and
$[\varphi(x), \partial_{x'}\vartheta(x')] = 2\pi i\delta(x - x')$. The parameters $v$ and $K$ are the velocity and Luttinger parameter of the Gaussian theory. The coefficients $g_2$ and $g_4$ describe the leading angular anisotropies, $g_{3\varphi}$ is the fugacity of the leading allowed dual vortex, and the total derivative proportional to $h$ shifts the characteristic momenta, which is the
continuum counterpart of the displaced doublet minima in Eq.~\eqref{eq:doublet_momentum_shift}.
The internal $\mathbb{Z}_2 \times \mathbb{Z}_2$ symmetry restricts the angular potential to harmonics $\cos(2n\vartheta)$.
One-site translation acts on the dual field as
$\varphi \rightarrow \varphi + 4\pi/3$ (see Appendix~\ref{app:integrating_singlet}), so that a vertex $e^{im\varphi}$ is invariant only when $m$ is a multiple of three. Consequently, $\cos\varphi$ and $\cos(2\varphi)$ are forbidden, while $\cos(3\varphi)$ is the leading allowed dual perturbation, followed by the higher harmonics $\cos(3n\varphi)$ with $n \geq 2$.
The microscopic weak-coupling construction underlying Eq.~\eqref{eq:sine_gordon}, valid near $(\lambda, V_{nn}) = (1, 0)$, is detailed in Appendix~\ref{app:weak_coupling_derivation}, particularly in \ref{app:quartic_reduction} through \ref{app:integrating_singlet}. There, the interacting Majorana doublet is bosonized, and the energy density of the massive singlet is replaced by its expectation value through mean-field decoupling.

In the convention of Eq.~\eqref{eq:sine_gordon}, the scaling dimensions of these perturbations are
\begin{align}
\Delta[\cos(2n\vartheta)] &= \frac{n^2}{K},&
\Delta[\cos(3 n \varphi)] &= 9 n^2 K.
\label{eq:vertex_scaling_dimensions}
\end{align}
For $1/2 < K < 2$, $\cos(2\vartheta)$ is relevant, whereas $\cos(4\vartheta)$, all higher angular harmonics, and all $\cos(3 n \varphi)$ are irrelevant. 
In this regime, for $g_2 > 0$, $\vartheta$ is pinned at $0$ or $\pi$, producing the $x$-FM phase, while for $g_2 < 0$, it is pinned at $\pi/2$ or $3\pi/2$, producing the $y$-FM phase. Along $g_2 = 0$, the remaining perturbations are irrelevant, leaving an extended Gaussian critical line with $c = 1$ and emergent $O(2)$ symmetry under $\vartheta \rightarrow \vartheta + \alpha$ and $\vartheta \rightarrow -\vartheta$. This prediction agrees with the $c = 1$ critical line obtained numerically in Sec.~\ref{sec:weak_nn}.

The identification $m_x \sim \langle\cos\vartheta\rangle$ and $m_y \sim \langle\sin\vartheta\rangle$
used above is, however, incomplete. The microscopic magnetizations carry Jordan-Wigner strings that run over all three Majorana flavors,
so that their continuum representatives involve the singlet Majorana as well as the doublet. As
argued in Appendix~\ref{app:integrating_singlet}, the singlet enters through the disorder field
$\mu_s$ of its Ising sector,
\begin{align}
m_x \sim \langle\mu_s\cos\vartheta\rangle,
\qquad
m_y \sim \langle\mu_s\sin\vartheta\rangle.
\label{eq:order_parameters_singlet}
\end{align}
Along the $c = 1$ line and throughout both ferromagnetic phases the singlet is gapped and lies on
the disordered side of its own transition. There $\mu_s$ may be replaced by the non-zero
constant $\langle\mu_s\rangle$, and the critical behavior of both magnetizations is carried by
$\cos\vartheta$ and $\sin\vartheta$ alone, as assumed above. The situation is different below
the $c = 1/2$ line (Figs.~\ref{fig:schem}(a) and \ref{fig:phase_nn}(a)), 
where the singlet sector is ordered. There $\langle\mu_s\rangle = 0$
and both magnetizations vanish even though $\vartheta$ is still pinned by $g_2$. That is why
neither the cluster nor the disordered phase carries local ferromagnetic order. 
The $c = 1/2$ line thus switches the singlet prefactor on and off, while the sign of $g_2$, which
changes across the $c = 1$ line, selects between the two orders.

Equation~\eqref{eq:sine_gordon} also constrains the critical exponents along the $c = 1$ DQC line.
The Luttinger parameter $K$ is non-universal, and the microscopic calculation fixes it only to leading order (see Appendix~\ref{app:bosonization}),
\begin{align}
K = 1 - \frac{V_{nn}}{\pi}
+\mathcal{O}(V_{nn}^2).
\label{eq:weak_coupling_K}
\end{align}
 No individual exponent can therefore be predicted quantitatively at finite $V_{nn}$. The exponents are, however, all functions of this single parameter, and hence not independent. The tuning operator $\cos(2\vartheta)$ has scaling dimension $1/K$, so its RG eigenvalue is $y_t = 2 - 1/K$ and the correlation-length exponent is
$\nu = 1/y_t = K/(2K - 1)$. 
With $\mu_s$ replaced by its expectation value, both order parameters therefore have scaling
dimension $\Delta[\cos\vartheta] = \Delta[\sin\vartheta] = 1/(4K)$, so that
$\beta = \nu\,\Delta[\cos\vartheta] = \nu/(4K) = 1/[4(2K - 1)]$. Eliminating $K$ between these two expressions leaves
\begin{align}
\beta = \frac{2\nu - 1}{4}.
\label{eq:beta_nu_relation}
\end{align}
At weak coupling $K \rightarrow 1$, so that $\nu \rightarrow 1$ and $\beta \rightarrow 1/4$. At finite $V_{nn}$, $K$ decreases and the two exponents drift together, tracing the one-parameter family fixed by Eq.~\eqref{eq:beta_nu_relation}. This is the hallmark of weak universality~\cite{suzuki_ptp_1974}: no single exponent is universal, but the relation between them is, and testing Eq.~\eqref{eq:beta_nu_relation} requires no knowledge of $K$.

Equation~\eqref{eq:beta_nu_relation} is exactly the relation obeyed along the critical line of the eight-vertex model, where Baxter's exact solution likewise yields exponents that vary continuously but remain locked to one another in this way~\cite{baxter_aop_1972, baxter_aop_1972b, kadanoff_aop_1979, baxter_book_1985}. This $\beta$-$\nu$ relation has been used to identify eight-vertex criticality in the interacting Kitaev chain~\cite{chepiga_prb_2023} and, in a setting closer to the present one, in a frustrated spin-1 chain with single-ion anisotropy~\cite{pronk_prb_2025}. In the latter, an $SU(2)_2$ multicritical point splits into Gaussian ($c = 1$) and Ising ($c = 1/2$) critical lines, with the Gaussian line directly separating phases that break incompatible $\mathbb{Z}_2$ symmetries. The $x$-FM to $y$-FM transition found here reproduces that structure: it too emerges from the splitting of an $SU(2)_2$ critical point, and it too directly connects two Landau-incompatible ordered phases. Together with the effective description of Eq.~\eqref{eq:sine_gordon}, this places the DQC line in the eight-vertex universality class. We test the identification directly below, extracting $\beta$ and $\nu$ independently and checking whether they satisfy Eq.~\eqref{eq:beta_nu_relation}.

To extract the critical exponents $\beta$ and $\nu$ across the $x$-FM $\leftrightarrow$ $y$-FM
phase transition, we use finite-size DMRG for open chains of different length $L$. However, a standard DMRG calculation 
for the Hamiltonian~\eqref{eq:H} 
is not ideal for extracting the 
order parameters, since on a finite chain the ground state is a
symmetric superposition of the two symmetry-broken states and both magnetizations vanish
identically. We can recover a non-zero signal by adding a weak symmetry-breaking term like
$\epsilon \sum_j \hat{\sigma}^x_j$ or $\epsilon \sum_j \hat{\sigma}^y_j$ with $\epsilon \ll 1$,
but the values we then obtain depend strongly on $\epsilon$ and are unreliable for a
finite-size scaling analysis. 
This is worst around the deconfined criticality, where the correlation length is large and the
enhanced $O(2)$ symmetry makes a weak field too soft to select a definite ordering direction.
To mitigate this issue, we instead select
the symmetry-broken states through the boundaries, and pin the leftmost spin along the
$x$-direction while the rightmost spin is pinned along $y$, by adding a \textit{strong} boundary pinning term
$-h_L \hat{\sigma}^x_1 - h_R \hat{\sigma}^y_L$ with $h_L, h_R \gg 1$, to the Hamiltonian~\eqref{eq:H}. 
By doing so, we force the
DMRG algorithm to produce symmetry-broken states in both phases. This does not spoil the
calculation, as the bulk Hamiltonian is untouched, and we measure the order parameters from the
middle of the chain, where the effect of the boundaries is smallest.

\begin{figure}[t]
    \centering
    \includegraphics[width=0.7\linewidth]{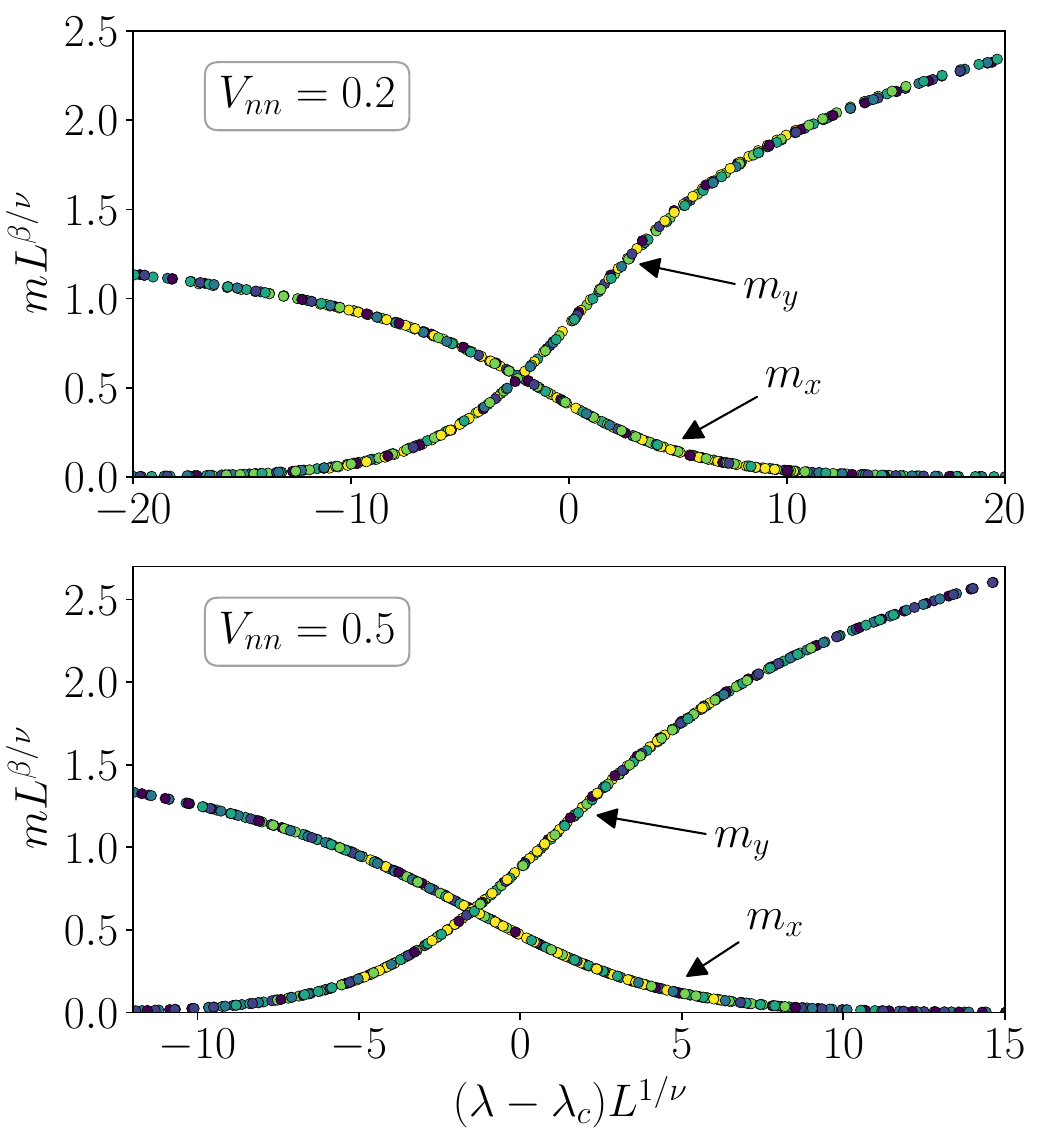}
    \caption{(Color online.) Finite-size scaling collapse of the magnetizations $m_x$ and $m_y$
    across the direct $x$-FM $\leftrightarrow$ $y$-FM transition, for $V_{nn} = 0.2$ (left) and
    $V_{nn} = 0.5$ (right). We plot $mL^{\beta/\nu}$ against $(\lambda - \lambda_c)L^{1/\nu}$,
    obtained using finite-size DMRG for open chains of length $L$ with pinned boundary spins,
    where different colors correspond to different $L \in [240, 840]$. The two magnetizations collapse onto
    complementary branches of a single curve, with $m_x$ decaying and $m_y$ growing as $\lambda$
    is increased through $\lambda_c$. The extracted $\lambda_c$, $\beta$ and $\nu$ are listed in
    Table~\ref{tab:exponents}.}
    \label{fig:fss}
\end{figure}

We use the standard critical finite-size scaling hypothesis
\begin{align}
m(\lambda) = L^{-\beta/\nu} f\left((\lambda - \lambda_c)L^{1/\nu} \right),
\label{eq:fss}
\end{align}
for a given $V_{nn}$, where $m$ is either $m_x$ or $m_y$ and $f(\cdot)$ is a non-universal scaling function. 
We treat $\lambda_c$, $\beta$ and $\nu$ as free parameters and determine them by
collapsing the data for all $L$ onto a single curve, and estimate the errors by statistical
bootstrapping as in Ref.~\cite{rakov_prb_2026}, where the data set is resampled many times and
the errors are taken as the standard deviation of the resulting distribution of fit parameters.
The collapses for $V_{nn} = 0.2$ and $0.5$ are shown in Fig.~\ref{fig:fss}, where $m_x$ and $m_y$
fall onto complementary branches of the same scaling function, and the extracted values are
listed in Table~\ref{tab:exponents}.

For both interaction strengths, $V_{nn} = 0.2$ and $0.5$, the fits to $m_x$ and $m_y$ return the same $\lambda_c$ within errors, confirming that the two orders vanish at a single point and ruling out a narrow intervening phase at the resolution of our system sizes. Both $\beta$ and $\nu$ grow with $V_{nn}$, which is the drift expected from the effective weak-coupling theory. Since $\nu = K/(2K - 1)$ and $\beta = 1/[4(2K - 1)]$ both decrease with $K$ and a growing exponent, therefore, means a decreasing Luttinger parameter, exactly as in Eq.~\eqref{eq:weak_coupling_K}. Inverting $\nu = K/(2K - 1)$ to read $K$ off the data gives $K = 0.98(3)$ at $V_{nn} = 0.2$ and $K = 0.89(2)$ at $V_{nn} = 0.5$, against the leading-order estimates $0.936$ and $0.841$. The agreement is good at both couplings, and it is not built in, since the weak-coupling theory is only leading order in $V_{nn}$ and makes no reference to the exponents.

\begin{table}[t]
\begin{ruledtabular}
\begin{tabular}{cccccc}
$V_{nn}$ & order parameter & $\lambda_c$ & $\beta$ & $\nu$ & $(2\nu - 1)/4$ \\
\colrule
0.2 & $m_x$ & 1.052(1)  & 0.26(2)  & 1.02(3) & 0.26(2) \\
    & $m_y$ & 1.051(1)  & 0.27(2)  & 1.02(3) & 0.26(2) \\
\colrule
0.5 & $m_x$ & 1.116(1)  & 0.32(2)  & 1.14(4) & 0.32(2) \\
    & $m_y$ & 1.116(1)  & 0.32(2)  & 1.15(4) & 0.33(2) \\
\end{tabular}
\end{ruledtabular}
\caption{Critical exponents at the direct $x$-FM to $y$-FM transition, extracted independently
from the two magnetizations by finite-size scaling. For each fit we also quote the extracted
critical point $\lambda_c$. The last column gives $(2\nu - 1)/4$, to be compared with the
measured $\beta$ as a test of Eq.~\eqref{eq:beta_nu_relation}.}
\label{tab:exponents}
\end{table}

The last column of Table~\ref{tab:exponents} tests the parameter-free relation of Eq.~\eqref{eq:beta_nu_relation}. For both interaction strengths and both order parameters, the independently extracted values of $\beta$ agree with $(2\nu - 1)/4$ within the numerical uncertainties, without any input from $K$. This agreement demonstrates that, although the individual exponents vary along the $x$-FM $\leftrightarrow$ $y$-FM critical line, their universal relation is preserved, identifying the DQC line with the eight-vertex weak universality class.

Finally, we should comment that the symmetry realization in Eq.~\eqref{eq:sine_gordon} is anomalous.  Pinning $\vartheta$ produces two degenerate vacua exchanged by $\mathcal{P}$, while pinning $\varphi$, were it possible, would produce three vacua cycled by one-site translation. The gradient term proportional to $h$ merely shifts the characteristic momenta and cannot open a gap. Consequently, the doublet sector cannot form a unique gapped ground state that preserves both $\mathcal{P}$ and translation: it must either remain gapless or spontaneously break one of these symmetries, a constraint familiar from the Lieb-Schultz-Mattis arguments~\cite{lieb_aop_1961, oshikawa_prl_2000}. This anomaly is emergent, since it constrains the low-energy doublet theory but is absent in the complete microscopic model, which also contains the singlet sector and admits symmetric gapped phases. In the weak-coupling regime, this anomaly is resolved either by the gapless DQC line, or by symmetry breaking with $\vartheta$ pinned on either side of $g_2=0$.

\section{Floating phase at stronger interactions}
\label{sec:strong_nn}

\begin{figure}
    \centering
    \includegraphics[width=\linewidth]{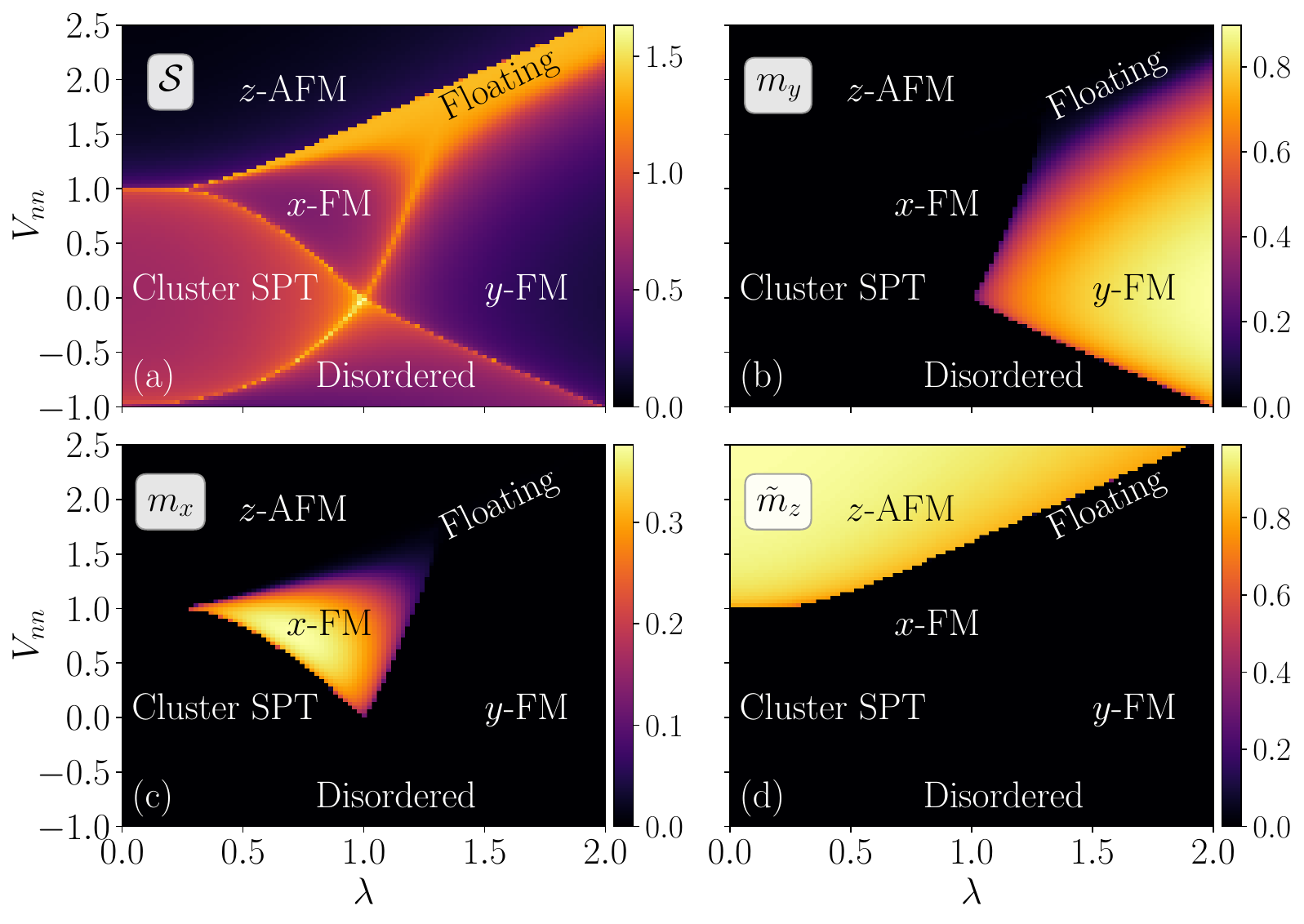}
    \caption{(Color online.) Phase diagram of the interacting CIM described by Eq.~\eqref{eq:H} over a
    wider range of parameters, obtained using VUMPS with bond dimension $\chi = 128$. We show the
    bipartite entanglement entropy $\mathcal{S}$ (a), the magnetizations $m_y$ (b) and $m_x$
    (c), and the staggered magnetization $\tilde{m}_z$ (d). For larger $V_{nn}$ the system
    breaks the one-site translation symmetry and stabilizes a phase with antiferromagnetic ordering along $z$,
    the $z$-AFM phase, which is separated 
    from the FM phases by a gapless floating phase. The $c = 1$ line separating the $x$-FM and $y$-FM phases 
    terminates where the floating phase opens.}
    \label{fig:phase_nn_large}
\end{figure}

So far we have treated $V_{nn}$ as a weak perturbation on top of the non-interacting CIM and worked in
the vicinity of $(\lambda, V_{nn}) = (1, 0)$. We now move away from that regime and follow the phase
diagram to larger interaction strengths.

The weak-coupling effective theory already indicates what to expect. According to
Eq.~\eqref{eq:weak_coupling_K}, increasing $V_{nn}$ decreases the Luttinger parameter, and once
$K < 1/2$ the tuning operator $\cos(2\vartheta)$ becomes irrelevant and no longer pins either
ferromagnetic order. The leading dual vertex $\cos(3\varphi)$ remains irrelevant as long as
$K > 2/9$, so within the window $2/9 < K < 1/2$ no symmetry-allowed perturbation of
Eq.~\eqref{eq:sine_gordon} survives, and a finite region of the phase diagram, rather than a
single line, flows to the Gaussian fixed point with $c = 1$.
The anomaly mentioned at the end of Sec.~\ref{sec:dqcl} is again resolved by gaplessness, but now over an extended region rather than along a single line, since neither field can be pinned and symmetry breaking is unavailable.

Such a region is not merely a widened critical line, because the total-derivative term in
Eq.~\eqref{eq:sine_gordon} remains. Completing the square in the Gaussian part gives the
background gradient
\begin{align}
\langle\partial_x\varphi\rangle = -\frac{2Kh}{v},
\label{eq:floating_background_gradient}
\end{align}
which varies continuously with the microscopic parameters and shifts the wavevector at which
correlations oscillate away from any commensurate value. The effective theory therefore predicts
\textit{a gapless phase} with $c = 1$ and algebraically decaying correlations at an incommensurate
wavevector, that is, a \textit{floating phase}~\cite{japaridze_jltp_1979,
pokrovsky_prl_1979}.

Figure~\ref{fig:phase_nn_large} shows the phase diagram over a wider range of $\lambda$ and
$V_{nn}$, and two new features appear at large $V_{nn}$. The first is a phase with
antiferromagnetic order along $z$, which we detect through the staggered magnetization
\begin{align}
\tilde{m}_z =  \frac{1}{L} \sum_j \langle (-1)^j \hat{\sigma}^z_j \rangle,
\label{eq:mz_stag}
\end{align}
shown in Fig.~\ref{fig:phase_nn_large}(d). We refer to this phase as the $z$-AFM phase. 
This  phase breaks the one-site translation
symmetry of $\hat{H}$ down to two-site translations, and is therefore of a different character
from all the phases encountered so far, none of which break translations. It also lies outside
the reach of Eq.~\eqref{eq:sine_gordon}, which was derived around $(\lambda, V_{nn}) = (1, 0)$ and
retains only the doublet sector.

The second is the floating phase, which occupies a wedge separating the $z$-AFM phase from both
ferromagnets. Its location is the one anticipated above. The $c = 1$
DQC line of Sec.~\ref{sec:dqcl} runs upward from $(\lambda, V_{nn}) = (1, 0)$, 
and terminates exactly where the wedge opens.
The critical line therefore does not simply end, but broadens into a phase, as expected once
$K$ falls below $1/2$.

\begin{figure}
    \centering
    \includegraphics[width=0.7\linewidth]{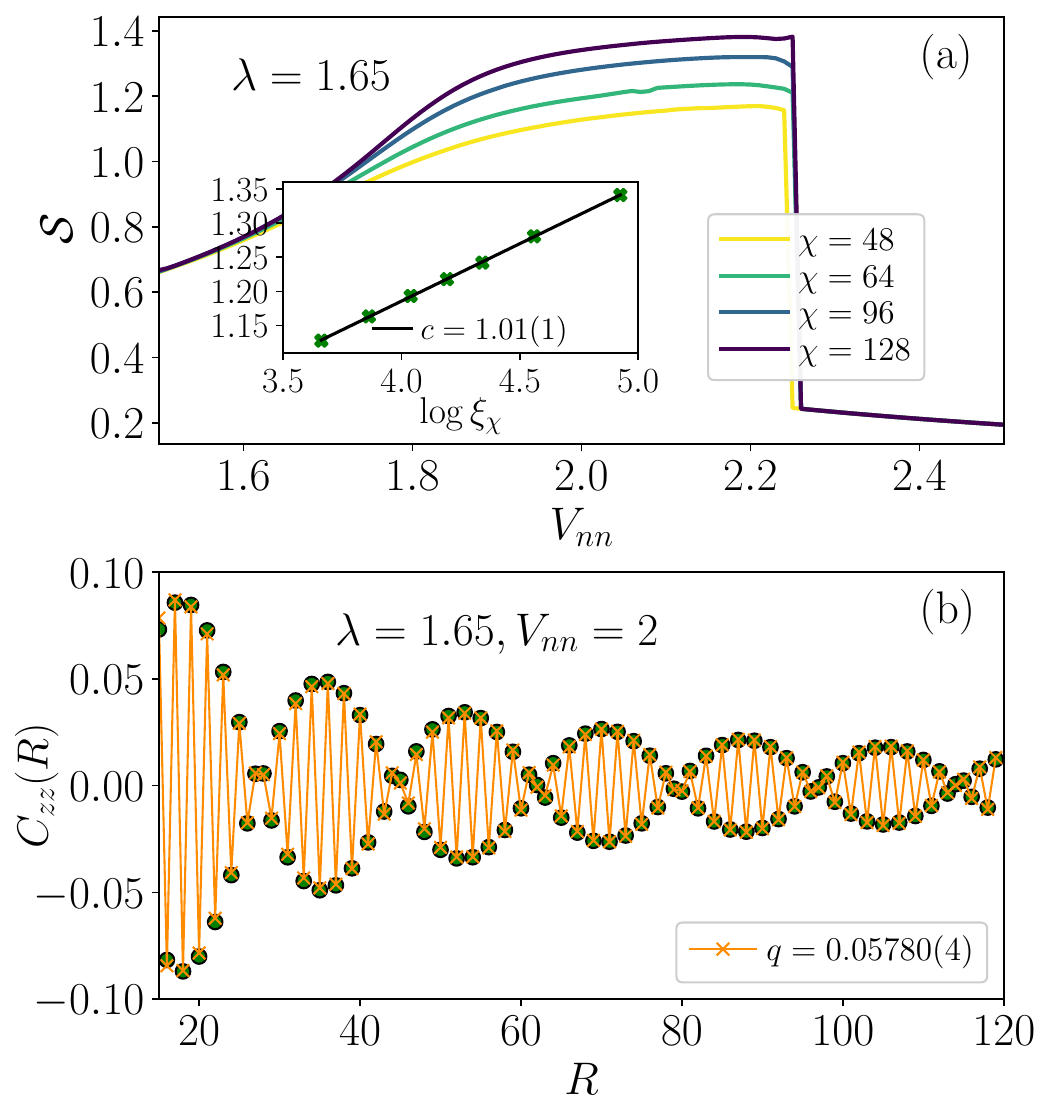}
    \caption{(Color online.) (a) Bipartite entanglement entropy across the floating phase as a
    function of $V_{nn}$ at fixed $\lambda = 1.65$, for different bond dimensions $\chi \in [48, 128]$, as the
    system passes from the $y$-FM phase through the floating phase into the $z$-AFM phase. The
    growth of $\mathcal{S}$ with $\chi$ inside the floating region signals a critical phase,
    whereas the abrupt drop at $V_{nn} \simeq 2.25$ marks the first-order transition into the
    $z$-AFM phase. (Inset) Finite-entanglement scaling of $\mathcal{S}$ at $\lambda = 1.65$ and
    $V_{nn} = 2$ according to Eq.~\eqref{eq:calabrese_cardy}, giving $c = 1.01(1)$. (b) Connected
    correlation function $C_{zz}(R)$ (Eq.~\eqref{eq:czz}) at the same point. The solid line is a
    fit to Eq.~\eqref{eq:czz_fit}, in which a commensurate staggered oscillation is modulated at
    the incommensurate wavevector $\pi q$ with $q = 0.05780(4)$ on top of an algebraically decaying envelope. 
    The wavevector depends
    smoothly on the parameters within the floating phase.}
    \label{fig:cuts_floating}
\end{figure}

We confirm the nature of the floating phase in Fig.~\ref{fig:cuts_floating}. Panel (a) shows the
bipartite entanglement entropy $\mathcal{S}$ along a cut at fixed $\lambda = 1.65$ as $V_{nn}$ is
increased from the $y$-FM phase to the $z$-AFM phase, through the floating phase. Within the
floating region $\mathcal{S}$ grows with $\chi$, as it must in a critical phase, and the
finite-entanglement scaling of Eq.~\eqref{eq:calabrese_cardy} at $\lambda = 1.65$ and $V_{nn} = 2$
gives $c = 1.01(1)$, shown in the inset, consistent with the single compact boson of
Eq.~\eqref{eq:sine_gordon}. At $V_{nn} \simeq 2.25$ the entropy drops abruptly to a
$\chi$-independent value, marking a first-order transition to the $z$-AFM phase. All transitions
involving the $z$-AFM phase are of this kind, with both $\tilde{m}_z$ and $\mathcal{S}$ changing
discontinuously. On the other hand, the transitions from the ferromagnetic phases into the floating phase occur
where $\cos(2\vartheta)$ becomes marginal, at $K = 1/2$, and are therefore of
BKT type~\cite{berezinsky_jetp_1970,kosterlitz_jpc_1973,kosterlitz_jpc_1974}.

Figure~\ref{fig:cuts_floating}(b) shows the connected correlation function
\begin{align}
C_{zz}(R) = \langle \hat{\sigma}^z_j \hat{\sigma}^z_{j + R}\rangle
          - \langle \hat{\sigma}^z_j\rangle\langle \hat{\sigma}^z_{j + R}\rangle
\label{eq:czz}
\end{align}
at the same point, together with a fit to
\begin{align}
C_{zz}(R) \simeq A\,(-1)^{R}\, R^{-\eta_{z}} \cos\left(\pi q R + \phi\right).
\label{eq:czz_fit}
\end{align}
The staggered factor $(-1)^R$ carries the commensurate antiferromagnetic oscillation at
wavevector $\pi$, with additional cosine modulations. Writing the product as a sum, the
correlations oscillate at the two wavevectors $\pi(1 \pm q)$, so that $q$ measures the detuning
from the commensurate $z$-AFM wavevector and $q \rightarrow 0$ recovers antiferromagnetic order.
The wavevector varies smoothly with $\lambda$ and $V_{nn}$ inside the phase and is generically incommensurate
with the lattice. This is the direct signature of the background gradient of
Eq.~\eqref{eq:floating_background_gradient}, and together with $c = 1$ it establishes the region
as a floating phase.
It should be noted that since the anomaly discussed in Sec.~\ref{sec:dqcl} originates in lattice translation rather than in an internal symmetry, the floating phase is not an intrinsically gapless topological phase~\cite{thorngren_prb_2021}, and we find no signature of protected edge modes in this phase.

\section{Conclusion}
\label{sec:conclu}

In this work, we have investigated the cluster Ising chain with nearest-neighbor interactions along the $z$ direction, combining tensor-network simulations with a weak-coupling field theory built from the Majorana representation of the lattice model and Abelian bosonization. In the non-interacting limit, the Jordan-Wigner transformation maps the model onto three identical, decoupled Majorana chains, revealing a hidden $O(3)$ flavor symmetry that rotates the three Majorana fields into one another. At the transition between the symmetry-protected topological cluster phase and the phase with ferromagnetic order along the $y$-direction (the $y$-FM phase), the common Majorana mass vanishes, and the resulting three massless Majorana modes realize the $SU(2)_2$ Wess-Zumino-Novikov-Witten theory with central charge $c = 3/2$. 

Our numerical phase diagram shows that interactions split this parent $c = 3/2$ transition into an Ising critical line with $c = 1/2$ and a Gaussian critical line with $c = 1$. Repulsive interactions stabilize an intervening phase with ferromagnetic order along the $x$-direction (the $x$-FM phase), whereas attractive interactions produce an intervening disordered phase and reverse the order of the two critical lines. We have explained this bifurcation microscopically by showing that the interaction breaks the hidden $O(3)$ symmetry while preserving a cyclic permutation symmetry $C_3$ among the three Majorana flavors. Under this residual symmetry, the Majorana triplet decomposes into a singlet and a real doublet. Closing the singlet gap leaves one massless Majorana mode and produces the Ising line, whereas closing the doublet gap leaves two massless Majorana modes, equivalently one Dirac fermion, and produces the Gaussian line. The leading-order weak-coupling predictions for the locations of these two lines are in good quantitative agreement with the numerical results.

The most striking consequence of this bifurcation has been the emergence of a direct $c = 1$ transition between the $x$-FM and $y$-FM phases for repulsive interactions. The $x$-FM phase breaks spin parity but preserves complex conjugation, whereas the $y$-FM phase breaks both symmetries separately while preserving their product. Since neither residual symmetry group is a subgroup of the other, these phases are Landau-incompatible and cannot be connected through a conventional continuous transition driven by either local order parameter. Nevertheless, we have found that both magnetizations vanish continuously at a common critical point, while finite-entanglement scaling has established a central charge $c = 1$. By bosonizing the critical Majorana doublet, we have obtained a compact-boson theory with an emergent $O(2)$ symmetry at criticality that continuously rotates the two ferromagnetic orders into one another. This theory has predicted continuously varying order-parameter and correlation-length exponents, $\beta$ and $\nu$, governed by the Luttinger parameter. We have extracted these exponents independently at two representative interaction strengths and have found that they satisfy the parameter-free relation $\beta = (2\nu - 1)/4$ for both magnetizations. Together, these results have identified the direct transition between the two Landau-incompatible phases as a deconfined quantum critical line in the eight-vertex weak universality class.

At stronger repulsive interactions, we have found that the phase diagram develops a $z$-AFM phase, characterized by staggered antiferromagnetic order along the $z$-direction, which spontaneously breaks lattice translation, and an intervening gapless floating phase separating it from both ferromagnets. Finite-entanglement scaling has established a central charge $c = 1$ within the floating phase, while the $z$-spin correlations  display algebraic decay modulated by an incommensurate wavevector that varies continuously with the microscopic parameters. The bosonized theory provides a natural explanation for both features of the floating phase: increasing the repulsive interaction reduces the Luttinger parameter until the anisotropy selecting between the two ferromagnetic orders becomes irrelevant, allowing the deconfined critical line to open into an extended gapless region, while a gradient term shifts the wavevector away from its commensurate value and produces the incommensurate correlations. The same theory has predicted Berezinskii-Kosterlitz-Thouless transitions between the ferromagnetic and floating phases. By contrast, our numerical results have shown that the transitions into the $z$-AFM phase are first order.

In summary, the interacting cluster Ising chain offers a simple microscopic setting in which a parent critical point with $c = 3/2$ resolves into an Ising and a Gaussian critical line, the latter realizing a deconfined critical line that opens into an incommensurate floating phase at stronger repulsion. What drives this is the reduction of an exact continuous flavor symmetry to a discrete subgroup, and the resulting critical lines along with the intervening phases are fixed not by accident but by the subgroup that the interaction spares. The Majorana and bosonization treatment developed here applies equally to other parent critical points whose central charge originates in a degeneracy enforced by a continuous internal symmetry, and should make the effect of interactions on them similarly predictable.

\acknowledgements
We thank Abhishodh Prakash for useful discussions and insightful comments. Sourabh and T.C. acknowledges the support from the Mphasis F1 Foundation via the Centre for Quantum Information, Communication, and Computing (CQuICC) at IIT Madras. 
B.B. acknowledges funding from the Wenner-Gren Foundations (project number UPD2024-0111).
M.T. acknowledges funding by the Deutsche Forschungsgemeinschaft (DFG, German Research Foundation) under Projektnummer 277101999 – TRR 183, and under Germany’s Excellence Strategy – Cluster of Excellence Matter and Light for Quantum Computing (ML4Q) EXC 2004/2 – 390534769.
T.C. acknowledges the support by the Young Faculty Initiation Grant (NFIG) at IIT Madras (Project No. RF24250775PHNFIG009162) and from the Anusandhan National Research Foundation (ANRF), India via the Prime Minister Early Career Research Grant ANRF/ECRG/2024/001198/PMS.
We thank National Supercomputing Mission (NSM) for providing computing resources of ‘PARAM RUDRA’ at the P. G. Senapathy Center For Computer Resources, Play Field Ave, Indian Institute of Technology Madras, Tamil Nadu 600036, which is implemented by C-DAC and supported by the Ministry of Electronics and Information Technology (MeitY) and Department of Science and Technology (DST), Government of India.
The tensor-network calculations have been performed using ITensors.jl~\cite{ITensor2022, ITensor2022codebase} and TenNetLib.jl~\cite{TenNetLib} libraries.

\appendix
\section{Microscopic derivation of the weak-coupling field theory}
\label{app:weak_coupling_derivation}

Here we derive the weak-coupling continuum theory used in the main text directly from the lattice Hamiltonian near $(\lambda, V_{nn}) = (1, 0)$, with $|\lambda - 1| \ll 1$ and $|V_{nn}| \ll 1$. 
All results below are accordingly obtained to first order in
$V_{nn}$, with $\mathcal{O}(V_{nn}^2)$ corrections consistently dropped.

We first establish the continuum normalization and obtain the interaction-induced splitting of the Majorana triplet into singlet and doublet sectors, including their excitation spectra. We then bosonize the interacting doublet sector to derive the effective compact-boson theory discussed in the main text. Throughout this Appendix, $a_0$ denotes the microscopic lattice spacing and $\ell = 3a_0$ the spacing of the three-site unit cell. The lattice operators are denoted by hats on top, whereas their continuum counterparts, i.e., the corresponding fields, are written without hats.

\subsection{Free lattice theory and continuum normalization}
\label{app:free_continuum}

We first determine the exact lattice dispersion and fix the continuum normalization used below. Reordering the second term of Eq.~\eqref{eq:H0_majorana} using
$\hat{\zeta}_j\hat{\eta}_{j + 1} = -\hat{\eta}_{j + 1}\hat{\zeta}_j$
and relabeling the site index gives
\begin{align}
\hat{H}_0
= i\sum_j\hat{\eta}_j
\left(\hat{\zeta}_{j + 2}
-\lambda\hat{\zeta}_{j - 1}\right).
\label{eq:app_free_eta_zeta}
\end{align}
We introduce the Fourier transforms
\begin{align}
\hat{\eta}_j &= \frac{1}{\sqrt{N}}\sum_p
e^{ipja_0}\hat{\eta}_p,&
\hat{\zeta}_j &= \frac{1}{\sqrt{N}}\sum_p
e^{ipja_0}\hat{\zeta}_p,
\label{eq:app_lattice_fourier}
\end{align}
where $N$ is the number of lattice sites and $p$ lies in the first Brillouin zone. Hermiticity of the real-space Majorana operators implies
$\hat{\eta}_p^\dagger = \hat{\eta}_{-p}$ and
$\hat{\zeta}_p^\dagger = \hat{\zeta}_{-p}$. In momentum space, Eq.~\eqref{eq:app_free_eta_zeta} becomes
\begin{align}
\hat{H}_0 = i\sum_p
\hat{\eta}_{-p}A_0(p)\hat{\zeta}_p,
\end{align}
where
\begin{align}
A_0(p)
= e^{2ipa_0} - \lambda e^{-ipa_0}
= e^{-ipa_0}\left(e^{3ipa_0} - \lambda\right).
\label{eq:app_free_lattice_amplitude}
\end{align}
With the lattice normalization
$\{\hat{\eta}_j, \hat{\eta}_{j'}\}
= \{\hat{\zeta}_j, \hat{\zeta}_{j'}\}
= 2\delta_{jj'}$ and
$\{\hat{\eta}_j, \hat{\zeta}_{j'}\} = 0$,
the positive excitation energy is
\begin{align}
E_0(p) = 2|A_0(p)|
= 2\sqrt{1 + \lambda^2 - 2\lambda\cos(3pa_0)}.
\label{eq:app_exact_free_dispersion}
\end{align}
At $\lambda = 1$, the spectrum is gapless at the three momenta
\begin{align}
p_0 = 0,\qquad p_\pm = \pm Q,\qquad
Q = \frac{2\pi}{3a_0}.
\label{eq:app_three_critical_momenta}
\end{align}
Expanding about any of these points, $p = p_\mu + q$ with
$p_\mu \in \{p_0, p_+, p_-\}$, and retaining the leading terms in $q$ and $\lambda - 1$, gives
\begin{align}
E_0^2(p_\mu + q)
\approx 4\left[(\lambda - 1)^2 + 9a_0^2q^2\right].
\label{eq:app_free_cone_expansion}
\end{align}
Comparison with the relativistic dispersion
$E^2 = m_0^2 + v_0^2q^2$ gives $v_0 = 6a_0$ and
$|m_0| = 2|\lambda - 1|$. Matching the expansion of $A_0(p)$ to the mass convention in Eq.~\eqref{eq:continuum_free} fixes the sign, yielding
\begin{align}
v_0 = 6a_0,\qquad m_0 = 2(\lambda - 1).
\label{eq:app_free_velocity_mass}
\end{align}

The mode near $p_0$ is self-conjugate under the Majorana reality condition and therefore describes a single real Majorana field. The modes near $p_+$ and $p_-$ are exchanged by this condition and together describe two real Majorana fields, equivalently one Dirac field. Upon folding these momenta into the reduced Brillouin zone of the three-site unit cell, the three gapless modes become the three flavor fields of Eq.~\eqref{eq:H0_flavors}. Their decomposition into a cyclic singlet and doublet in the interacting theory is derived below in Appendix~\ref{app:singlet_doublet}.

For the continuum limit, we set $x = k\ell$ and introduce the fields
$\hat{\eta}_k^a = \sqrt{2\ell}\,\eta^a(x)$ and
$\hat{\zeta}_k^a = \sqrt{2\ell}\,\zeta^a(x)$. They satisfy
$\{\eta^a(x), \eta^b(x')\}
= \{\zeta^a(x), \zeta^b(x')\}
= \delta^{ab}\delta(x - x')$ and
$\{\eta^a(x), \zeta^b(x')\} = 0$.
Using $\sum_k \rightarrow \ell^{-1}\int dx$ and expanding
Eq.~\eqref{eq:H0_flavors} to first order in $\ell$ gives
\begin{align}
\hat{H}_0 \rightarrow
i\sum_{a = 1}^3\int dx\,
\eta^a\left(v_0\partial_x - m_0\right)\zeta^a,
\label{eq:app_free_continuum_hamiltonian}
\end{align}
where $v_0$ and $m_0$ are given in
Eq.~\eqref{eq:app_free_velocity_mass}. This reproduces the continuum Hamiltonian in Eq.~\eqref{eq:continuum_free}.

\subsection{Lattice interaction and normal ordering}
\label{app:normal_ordering}

Using $\hat{\sigma}^z_j = -i\hat{\zeta}_j\hat{\eta}_j$ and the flavor decomposition in Eq.~\eqref{eq:flavor_define}, 
the interaction part of the Hamiltonian~\eqref{eq:H} becomes
\begin{align}
\hat{H}_V = -V_{nn}\sum_k\big[{}&
(\hat{\zeta}^2_k\hat{\eta}^2_k)
(\hat{\eta}^1_k\hat{\zeta}^3_k)
+ (\hat{\zeta}^3_k\hat{\eta}^3_k)
(\hat{\eta}^2_k\hat{\zeta}^1_{k + 1}) \notag\\
&+ (\hat{\zeta}^1_{k + 1}\hat{\eta}^1_{k + 1})
(\hat{\eta}^3_k\hat{\zeta}^2_{k + 1})\big].
\label{eq:app_three_bonds}
\end{align}
We perform the weak-coupling expansion about $(\lambda, V_{nn}) = (1, 0)$ by normal ordering with respect to the free critical ground state, whose expectation values are denoted by $\langle\cdots\rangle_0$. Since $\hat{H}_0$ does not mix the three flavors, all interflavor contractions vanish. The only non-zero contraction within each term of Eq.~\eqref{eq:app_three_bonds} is therefore the same-flavor intracell contraction
\begin{align}
\left\langle i\hat{\zeta}^a_k\hat{\eta}^a_k\right\rangle_0
&= -\int_{-\pi}^{\pi}\frac{dq}{2\pi}
\frac{1 - \cos q}{\sqrt{2 - 2\cos q}}
= -\frac{2}{\pi},
\label{eq:app_critical_contraction}
\end{align}
where $q$ is the dimensionless momentum conjugate to the unit-cell index $k$. 

Separating the interaction into its contracted quadratic part and fully normal-ordered quartic remainder,
$\hat{H}_V = \hat{H}_V^{(2)} + \hat{H}_V^{(4)}$, gives
\begin{align}
\hat{H}_V^{(2)} = -\frac{2iV_{nn}}{\pi}\sum_k\big(
\hat{\eta}_k^1\hat{\zeta}_k^3
+ \hat{\eta}_k^2\hat{\zeta}_{k + 1}^1
+ \hat{\eta}_k^3\hat{\zeta}_{k + 1}^2\big).
\label{eq:app_lattice_self_energy}
\end{align}
This quadratic contribution produces the leading singlet-doublet mass splitting. The remaining term $\hat{H}_V^{(4)}$ is the fully normal-ordered form of Eq.~\eqref{eq:app_three_bonds} -- it generates the marginal interactions retained in the bosonized continuum theory derived in Appendix~\ref{app:quartic_reduction}.

Using the continuum normalization established above, we write $\boldsymbol{\eta} = (\eta^1, \eta^2, \eta^3)^T$ and $\boldsymbol{\zeta} = (\zeta^1, \zeta^2, \zeta^3)^T$. The continuum limit of Eq.~\eqref{eq:app_lattice_self_energy} is
\begin{align}
H_V^{(2)}
&= -\frac{4iV_{nn}}{\pi}\int dx\,
\boldsymbol{\eta}^{T}C^2\boldsymbol{\zeta} \notag\\
- &\frac{4i\ell V_{nn}}{\pi}\int dx\,
\left(\eta^2\partial_x\zeta^1
+ \eta^3\partial_x\zeta^2\right) + \cdots .
\label{eq:app_continuum_self_energy}
\end{align}
Here $C$ is the cyclic matrix defined in Eq.~\eqref{eq:cyclic_flavor_matrix}. The first term is the $O(3)$-breaking mass correction responsible for the bifurcation. The derivative terms vanish at zero momentum and therefore do not affect the mass matrix or the critical lines to this order, so we omit them from the following mass analysis.

\subsection{Cyclic symmetry and singlet-doublet splitting}
\label{app:singlet_doublet}

The quadratic correction in Eq.~\eqref{eq:app_continuum_self_energy} breaks the $O(3)$ flavor symmetry but preserves the cyclic subgroup $C_3$. To determine the resulting mass structure, we write the renormalized zero-momentum mass term as
\begin{align}
H_M = -i\int dx\,
\boldsymbol{\eta}^{T}M\boldsymbol{\zeta}.
\label{eq:app_mass_term_definition}
\end{align}
Hermiticity requires the $3 \times 3$ mass matrix $M$ to be real, but does not require it to be symmetric. Cyclic invariance imposes
$C^T M C = M$, or equivalently $MC = CM$. Since $C^3 = \mathbb{I}_3$, the real matrices commuting with $C$ are spanned by $\mathbb{I}_3$, $C$, and $C^2$. Separating the symmetric and antisymmetric combinations gives
\begin{align}
M = \widetilde m_0\mathbb{I}_3
+ m_1(C + C^2) + m_2(C - C^2),
\label{eq:app_general_mass_matrix}
\end{align}
where $\widetilde m_0$ is the renormalized flavor-independent mass, while $m_1$ and $m_2$ are the symmetric and antisymmetric flavor-mixing coefficients, respectively. 

We introduce the orthonormal flavor basis
\begin{align}
\boldsymbol{e}_s &= \frac{1}{\sqrt{3}}(1, 1, 1)^T, \notag\\
\boldsymbol{e}_1 &= \frac{1}{\sqrt{6}}(2, -1, -1)^T,\qquad
\boldsymbol{e}_2 = \frac{1}{\sqrt{2}}(0, 1, -1)^T,
\label{eq:app_flavor_basis}
\end{align}
and the orthogonal matrix
$U = (\boldsymbol{e}_s, \boldsymbol{e}_1, \boldsymbol{e}_2)$. The symmetric vector $\boldsymbol{e}_s$ is invariant under $C$, whereas $\boldsymbol{e}_1$ and $\boldsymbol{e}_2$ span the invariant plane on which $C$ acts as a rotation by $-2\pi/3$. For either $\boldsymbol{f} = \boldsymbol{\eta}$ or $\boldsymbol{\zeta}$, the singlet and doublet fields are
\begin{align}
f_s = \boldsymbol{e}_s^T\boldsymbol{f}, \quad \boldsymbol{f}_d = (\boldsymbol{e}_1^T\boldsymbol{f},
\boldsymbol{e}_2^T\boldsymbol{f})^T,
\label{eq:app_flavor_transformation}
\end{align}
respectively. In this basis, the mass matrix becomes
\begin{align}
U^TMU =
\begin{pmatrix}
m_s & 0 & 0\\
0 & m_d & \sqrt{3}\,m_2\\
0 & -\sqrt{3}\,m_2 & m_d
\end{pmatrix},
\label{eq:app_mass_block_form}
\end{align}
where
\begin{align}
m_s &= \widetilde m_0 + 2m_1,\qquad
m_d = \widetilde m_0 - m_1.
\label{eq:app_singlet_doublet_masses}
\end{align}
Thus, the residual cyclic symmetry allows the singlet and doublet to acquire distinct gap-opening masses, while $m_2$ produces an antisymmetric mixing within the doublet.

We now match the symmetry-allowed mass matrix to the microscopic interaction. Combining the free mass in Eq.~\eqref{eq:app_free_velocity_mass} with the nonderivative contribution in Eq.~\eqref{eq:app_continuum_self_energy} gives
\begin{align}
M = 2(\lambda - 1)\mathbb{I}_3
+ \frac{4V_{nn}}{\pi}C^2
+\mathcal{O}(V_{nn}^2).
\label{eq:app_matched_mass_matrix}
\end{align}
Using
$C^2 = [(C + C^2) - (C - C^2)]/2$
and comparing with Eq.~\eqref{eq:app_general_mass_matrix}, we find
\begin{align}
\widetilde m_0 &= 2(\lambda - 1)
+\mathcal{O}(V_{nn}^2), \notag\\
m_1 &= \frac{2V_{nn}}{\pi}
+\mathcal{O}(V_{nn}^2), \notag\\
m_2 &= -\frac{2V_{nn}}{\pi}
+\mathcal{O}(V_{nn}^2).
\label{eq:app_matching_coefficients}
\end{align}
Substitution into Eq.~\eqref{eq:app_singlet_doublet_masses} reproduces the singlet and doublet masses quoted in Eq.~\eqref{eq:split_masses_linear}. Their zeros determine the critical lines
\begin{align}
\lambda_s &= 1 - \frac{2V_{nn}}{\pi}
+\mathcal{O}(V_{nn}^2),&
\lambda_d &= 1 + \frac{V_{nn}}{\pi}
+\mathcal{O}(V_{nn}^2).
\label{eq:app_critical_lines}
\end{align}
The two lines meet at $(\lambda, V_{nn}) = (1, 0)$ and exchange their order under $V_{nn} \rightarrow -V_{nn}$.

\subsection{Singlet and doublet dispersions}
\label{app:dispersion_derivation}

We now derive the dispersions quoted in Eq.~\eqref{eq:singlet_doublet_dispersions} and determine the role of the antisymmetric coefficient $m_2$. The derivative corrections in Eq.~\eqref{eq:app_continuum_self_energy} vanish at zero momentum and therefore do not affect the gaps. Since the displaced doublet minima occur at momenta of order $V_{nn}$, these corrections modify their positions only at order $V_{nn}^2$. To the order retained here, we neglect these derivative corrections when determining the gaps and momentum shifts, and allow for renormalized sector velocities $v_s$ and $v_d$. 

Introducing
\begin{align}
\mathcal{J} =
\begin{pmatrix}
0 & 1\\
-1 & 0
\end{pmatrix},
\qquad
\mathcal{J}^2 = -\mathbb{I}_2,
\end{align}
the quadratic Hamiltonians are
\begin{align}
H_s ={}& i\int dx\,
\eta_s\left(v_s\partial_x - m_s\right)\zeta_s, \notag\\
H_d ={}& i\int dx\,
\boldsymbol{\eta}_d^T
\left[
v_d\partial_x\mathbb{I}_2
- m_d\mathbb{I}_2
- \sqrt{3}\,m_2\mathcal{J}
\right]\boldsymbol{\zeta}_d.
\label{eq:app_sector_hamiltonians}
\end{align}
We use the Fourier convention
\begin{align}
f_\mu(x) &= \int_k e^{ikx}f_{\mu, k},&
\int_k &\equiv \int_{-\infty}^{\infty}\frac{dk}{2\pi},
\label{eq:app_continuum_fourier}
\end{align}
where $f = \eta, \zeta$ and $\mu = s, 1, 2$. Hermiticity of the Majorana fields implies
\begin{align}
\eta_{\mu, k}^\dagger = \eta_{\mu, -k},\qquad
\zeta_{\mu, k}^\dagger = \zeta_{\mu, -k}.
\label{eq:app_momentum_majorana_reality}
\end{align}
Introducing the Majorana-Nambu columns
\begin{align}
\Psi_{s, k} &= (\eta_{s, k}, \zeta_{s, k})^T, \notag\\
\Psi_{d, k} &= (\eta_{1, k}, \eta_{2, k},
\zeta_{1, k}, \zeta_{2, k})^T,
\label{eq:app_nambu_columns}
\end{align}
the quadratic Hamiltonian for either sector $\nu = s, d$ becomes
\begin{align}
H_\nu &= \frac{1}{2}\int_k
\Psi_{\nu, k}^\dagger h_\nu(k)\Psi_{\nu, k}, \notag\\
h_\nu(k) &=
\begin{pmatrix}
0 & -i\mathsf{D}_\nu(k)\\
i\mathsf{D}_\nu^\dagger(k) & 0
\end{pmatrix},
\label{eq:app_single_particle_hamiltonian}
\end{align}
where the factor $1/2$ removes the double counting between $k$ and $-k$. The momentum-space coupling matrices are
\begin{align}
\mathsf{D}_s(k) &= m_s - iv_sk, \notag\\
\mathsf{D}_d(k) &=
(m_d - iv_dk)\mathbb{I}_2
+ \sqrt{3}\,m_2\mathcal{J}.
\label{eq:app_momentum_couplings}
\end{align}
The resulting positive-energy dispersions are
\begin{align}
E_s^2(k) &= m_s^2 + v_s^2k^2, \notag\\
E_{d, \pm}^2(k) &=
m_d^2 + \left(v_dk \pm \sqrt{3}\,m_2\right)^2,
\label{eq:app_doublet_dispersion}
\end{align}
which reproduce Eq.~\eqref{eq:singlet_doublet_dispersions} of the main text. The two doublet branches satisfy
$E_{d, +}(k) = E_{d, -}(-k)$ and have minima at
\begin{align}
q_\pm = \mp \frac{\sqrt{3}\,m_2}{v_d},
\qquad
E_{d, \pm}(q_\pm) = |m_d|.
\label{eq:app_doublet_minima}
\end{align}
The momenta $q_\pm$ reproduce Eq.~\eqref{eq:doublet_momentum_shift}. Thus, $m_2$ shifts the doublet minima to opposite momenta without changing their common gap. Consequently, $m_s = 0$ produces one massless Majorana mode with $c = 1/2$, whereas $m_d = 0$ produces a massless Majorana doublet, equivalently one Dirac mode, with $c = 1$, as used in the interpretation of the critical bifurcation in Sec.~\ref{sec:weak_nn}.

\subsection{Interacting singlet-doublet continuum theory}
\label{app:quartic_reduction}

The preceding analysis determines the interaction-induced masses and dispersions from the quadratic contribution to the normal-ordered interaction. We now retain the residual quartic term $\hat{H}_V^{(4)}$ introduced below Eq.~\eqref{eq:app_lattice_self_energy} to obtain the interacting low-energy theory. To leading order in the continuum expansion, the cell displacements in Eq.~\eqref{eq:app_three_bonds} may be neglected, as they contribute only derivative corrections. For $f = \eta, \zeta$, we use superscripts $f^a$, with $a = 1, 2, 3$, for the original flavor fields and subscripts $f_\mu$, with $\mu = s, 1, 2$, for their singlet and doublet components. By inverting the relations of Eq.~\eqref{eq:app_flavor_transformation}, the original fields are expressed in the singlet-doublet basis as
\begin{align}
\zeta^1 &= \frac{\zeta_s}{\sqrt{3}} + \sqrt{\frac{2}{3}}\,\zeta_1, \notag\\
\zeta^2 &= \frac{\zeta_s}{\sqrt{3}} - \frac{\zeta_1}{\sqrt{6}}
+ \frac{\zeta_2}{\sqrt{2}}, \notag\\
\zeta^3 &= \frac{\zeta_s}{\sqrt{3}} - \frac{\zeta_1}{\sqrt{6}}
- \frac{\zeta_2}{\sqrt{2}},
\label{eq:app_inverse_flavor_transformation}
\end{align}
and similarly for the $\eta$ fields.

For each sector $\mu = s, 1, 2$, we introduce left- and right-moving Majorana fields through
\begin{align}
\eta_\mu = \frac{\xi_L^\mu + \xi_R^\mu}{\sqrt{2}},
\qquad
\zeta_\mu = \frac{\xi_L^\mu - \xi_R^\mu}{\sqrt{2}}.
\label{eq:app_chiral_majoranas}
\end{align}
For the chiral fields $\xi_{L, R}^\mu$, the sector label is written as a superscript because the subscript labels chirality. The diagonal kinetic and gap-opening mass bilinears in Eq.~\eqref{eq:app_sector_hamiltonians} then become
\begin{align}
i\eta_\mu v_\mu\partial_x\zeta_\mu
&= \frac{iv_\mu}{2}
\left(\xi_L^\mu\partial_x\xi_L^\mu
-\xi_R^\mu\partial_x\xi_R^\mu\right), \notag\\
-i\eta_\mu\zeta_\mu
&= i\xi_L^\mu\xi_R^\mu,
\label{eq:app_chiral_quadratic_terms}
\end{align}
up to a total derivative in the first line, with $v_\mu = v_s$ for the singlet and $v_\mu = v_d$ for the two doublet components.

Substituting Eqs.~\eqref{eq:app_inverse_flavor_transformation} and \eqref{eq:app_chiral_majoranas} into the normal-ordered interaction, including the overall minus sign in $\hat{H}_V$, and collecting identical terms gives
\begin{align}
H_V^{(4)} = g\int dx\,\mathcal{V}_4,
\qquad
g = 4\ell V_{nn} = 12a_0V_{nn},
\label{eq:app_quartic_coupling}
\end{align}
where $\mathcal{V}_4$ is the continuum quartic interaction density,
\begin{align}
\mathcal{V}_4 ={}&
\frac{\sqrt{3}}{2}\xi_L^s\xi_R^s
\left(\xi_L^1\xi_L^2 - \xi_R^1\xi_R^2\right) \notag\\
&+\frac{1}{2}\xi_L^s\xi_R^s
\left(\xi_L^1\xi_R^1 + \xi_L^2\xi_R^2\right)
- \xi_L^1\xi_R^1\xi_L^2\xi_R^2.
\label{eq:app_quartic_majorana_form}
\end{align}
Normal ordering is implicit. The factor $4\ell$ follows from the continuum normalization in Appendix~\ref{app:free_continuum}: the four Majorana fields contribute $(2\ell)^2$, while the unit-cell sum contributes $\ell^{-1}$.

The two real components of the doublet are combined into the chiral components of a Dirac fermion,
\begin{align}
\psi_L = \frac{\xi_L^1 - i\xi_L^2}{\sqrt{2}},
\qquad
\psi_R = \frac{\xi_R^1 + i\xi_R^2}{\sqrt{2}}.
\label{eq:app_dirac_definition}
\end{align}
With this relative sign convention, the common doublet mass is represented by a pairing bilinear. We introduce the singlet energy density $\varepsilon_s$, the normal-ordered chiral densities $\rho_L$ and $\rho_R$, and the doublet mass bilinear $\mathcal{D}$ as
\begin{align}
\varepsilon_s &= i\xi_L^s\xi_R^s, \notag\\
\rho_L &={:}\psi_L^\dagger\psi_L{:}
= -i\xi_L^1\xi_L^2, \notag\\
\rho_R &={:}\psi_R^\dagger\psi_R{:}
= i\xi_R^1\xi_R^2, \notag\\
\mathcal{D} &=
i\left(\xi_L^1\xi_R^1 + \xi_L^2\xi_R^2\right) \notag\\
&= i\left(\psi_L^\dagger\psi_R^\dagger + \psi_L\psi_R\right).
\label{eq:app_bilinear_definitions}
\end{align}
The antisymmetric doublet coupling in Eq.~\eqref{eq:app_sector_hamiltonians} becomes
\begin{align}
-i\sqrt{3}\,m_2
\left(\eta_1\zeta_2 - \eta_2\zeta_1\right)
&= -i\sqrt{3}\,m_2
\left(\xi_L^1\xi_L^2 - \xi_R^1\xi_R^2\right) \notag\\
&= \sqrt{3}\,m_2(\rho_L + \rho_R).
\label{eq:app_m2_density_coupling}
\end{align}
Thus, $m_2$ couples to the total Dirac density and produces the momentum shifts in Eq.~\eqref{eq:app_doublet_minima}.

The remaining quartic identities are
\begin{align}
-\xi_L^1\xi_R^1\xi_L^2\xi_R^2
&= \rho_L\rho_R, \notag\\
\xi_L^s\xi_R^s
\left(\xi_L^1\xi_R^1 + \xi_L^2\xi_R^2\right)
&= -\varepsilon_s\mathcal{D}, \notag\\
\xi_L^s\xi_R^s
\left(\xi_L^1\xi_L^2 - \xi_R^1\xi_R^2\right)
&= \varepsilon_s(\rho_L + \rho_R).
\label{eq:app_quartic_identities}
\end{align}
The quartic interaction in Eq.~\eqref{eq:app_quartic_majorana_form} therefore becomes
\begin{align}
H_V^{(4)} = g\int dx\left[
\rho_L\rho_R
- \frac{1}{2}\varepsilon_s\mathcal{D}
+ \frac{\sqrt{3}}{2}\varepsilon_s(\rho_L + \rho_R)
\right].
\label{eq:app_residual_vertex}
\end{align}

Finally, denoting the kinetic terms in this basis as
\begin{align}
H_{s, 0} &= \frac{iv_s}{2}\int dx\,
\left(\xi_L^s\partial_x\xi_L^s
- \xi_R^s\partial_x\xi_R^s\right), \notag\\
H_{d, 0} &= iv_d\int dx\,
\left(\psi_L^\dagger\partial_x\psi_L
- \psi_R^\dagger\partial_x\psi_R\right),
\label{eq:app_kinetic_hamiltonians}
\end{align}
the fermionic continuum Hamiltonian to the order retained is
\begin{align}
H ={}& H_{s, 0} + H_{d, 0}
+ \int dx\left(m_s\varepsilon_s + m_d\mathcal{D}\right) \notag\\
&+ \sqrt{3}\,m_2\int dx\,(\rho_L + \rho_R) \notag\\
&+ g\int dx\left[
\rho_L\rho_R - \frac{1}{2}\varepsilon_s\mathcal{D}\right] \notag\\
&+ \frac{\sqrt{3}\,g}{2}\int dx\,
\varepsilon_s(\rho_L + \rho_R) + \cdots .
\label{eq:app_complete_fermionic_theory}
\end{align}
Here, the ellipsis denotes derivative corrections and terms beyond the retained order in the weak-coupling expansion. This fermionic continuum Hamiltonian provides the starting point for bosonizing the interacting doublet sector.

\subsection{Bosonization of the interacting doublet}
\label{app:bosonization}

The doublet forms a single Dirac fermion, allowing its density-density interaction to be absorbed into a Gaussian bosonic theory~\cite{gogolin_2004, giamarchi_2003}. We introduce the compact fields $\vartheta \equiv \vartheta + 2\pi$ and $\varphi \equiv \varphi + 2\pi$, satisfying
\begin{align}
[\varphi(x), \partial_{x'}\vartheta(x')]
= 2\pi i\delta(x - x').
\label{eq:app_boson_commutator}
\end{align}
The normalization adopted here coincides with that of Giamarchi~\cite{giamarchi_2003} upon identifying $\varphi = 2\varphi_{\mathrm{Giamarchi}}$.
Writing $(\psi_{-1}, \kappa_{-1}) \equiv (\psi_L, \kappa_L)$ and
$(\psi_{+1}, \kappa_{+1}) \equiv (\psi_R, \kappa_R)$, we use the bosonization convention
\begin{align}
\psi_r(x) = \frac{\kappa_r}{\sqrt{2\pi\alpha_0}}
\exp\left[
-i\left(\frac{r}{2}\varphi(x) - \vartheta(x)\right)
\right].
\label{eq:app_bosonization_dictionary}
\end{align}
Here, $\alpha_0$ is a short-distance cutoff, and the Hermitian Klein factors $\kappa_r$ satisfy
$\{\kappa_r, \kappa_{r'}\} = 2\delta_{rr'}$. 
Using the short-distance operator-product expansion of the bosonized fermion fields, we get the normal-ordered chiral densities as
\begin{align}
\rho_L &= -\frac{1}{4\pi}\partial_x\varphi
- \frac{1}{2\pi}\partial_x\vartheta, \notag\\
\rho_R &= -\frac{1}{4\pi}\partial_x\varphi
+ \frac{1}{2\pi}\partial_x\vartheta,
\label{eq:app_chiral_densities}
\end{align}
and hence
\begin{align}
\rho_L + \rho_R
&= -\frac{1}{2\pi}\partial_x\varphi, \notag\\
\rho_L\rho_R
&= \frac{1}{16\pi^2}(\partial_x\varphi)^2
- \frac{1}{4\pi^2}(\partial_x\vartheta)^2.
\label{eq:app_density_dictionary}
\end{align}
Fixing the Klein-factor sector such that
$i\kappa_L\kappa_R = 1$, the doublet mass bilinear becomes
\begin{align}
\mathcal{D}
= \frac{1}{\pi\alpha_0}\cos(2\vartheta).
\label{eq:app_mass_bosonization}
\end{align}
The opposite choice of Klein-factor sector is equivalent to a constant shift of $\vartheta$ and does not affect the resulting theory.

The doublet kinetic term in Eq.~\eqref{eq:app_kinetic_hamiltonians} bosonizes as
\begin{align}
H_{d, 0}
= \frac{v_d}{2\pi}\int dx\left[
(\partial_x\vartheta)^2
+\frac{1}{4}(\partial_x\varphi)^2
\right].
\label{eq:app_free_dirac_bosonized}
\end{align}
Combining this term with the forward-scattering interaction
$g\int dx\,\rho_L\rho_R$ gives
\begin{align}
H_G = \frac{v}{2\pi}\int dx\left[
K(\partial_x\vartheta)^2
+\frac{1}{4K}(\partial_x\varphi)^2
\right],
\label{eq:app_gaussian_hamiltonian}
\end{align}
where
\begin{align}
vK &= v_d - \frac{g}{2\pi},&
\frac{v}{K} &= v_d + \frac{g}{2\pi}.
\label{eq:app_velocity_K_relations}
\end{align}
Therefore,
\begin{align}
v &= \sqrt{v_d^2 - \left(\frac{g}{2\pi}\right)^2}, \notag\\
K &= \sqrt{
\frac{v_d - g/(2\pi)}
{v_d + g/(2\pi)}
}.
\label{eq:app_velocity_K}
\end{align}
Using $v_d = 6a_0 + \mathcal{O}(V_{nn}a_0)$ and
$g = 12a_0V_{nn}$, we obtain
\begin{align}
K = 1 - \frac{V_{nn}}{\pi}
+\mathcal{O}(V_{nn}^2),
\label{eq:app_weak_coupling_K}
\end{align}
which reproduces Eq.~\eqref{eq:weak_coupling_K}. The linear correction to $v_d$ cancels from $K$ at this order.

Applying Eqs.~\eqref{eq:app_density_dictionary} and \eqref{eq:app_mass_bosonization} to Eq.~\eqref{eq:app_complete_fermionic_theory} gives
\begin{align}
H ={}& H_{s, 0}
+ m_s\int dx\,\varepsilon_s
+ H_G \notag\\
&+ \frac{1}{\pi\alpha_0}\int dx\,
\left(m_d - \frac{g}{2}\varepsilon_s\right)
\cos(2\vartheta) \notag\\
&- \frac{\sqrt{3}}{2\pi}\int dx\,
\left(m_2 + \frac{g}{2}\varepsilon_s\right)
\partial_x\varphi + \cdots .
\label{eq:app_coupled_singlet_doublet}
\end{align}
Equation~\eqref{eq:app_coupled_singlet_doublet} is the bosonized doublet theory coupled to the singlet energy density. 

\subsection{Effective compact-boson Hamiltonian}
\label{app:integrating_singlet}

Along the doublet transition, the singlet remains massive, with $m_s \neq 0$. At energies below the singlet gap, its energy density may be replaced to leading order by
\begin{align}
\varepsilon_s(x)
= \langle\varepsilon_s\rangle + \delta\varepsilon_s(x),
\label{eq:app_singlet_energy_decomposition}
\end{align}
where the fluctuations $\delta\varepsilon_s$ generate only local corrections to the low-energy
doublet theory. In Eq.~\eqref{eq:app_complete_fermionic_theory}, all couplings between the singlet and doublet sectors are mediated by the singlet energy density $\varepsilon_s$ and are proportional to $g$. The singlet remains gapped along the doublet transition and throughout both the adjacent phases. Therefore, dropping $\delta\varepsilon_s$ by mean-field decoupling
leads to corrections at the order of $g^2$, i.e., $\mathcal{O}(V_{nn}^2)$.
Retaining the leading contribution $\langle\varepsilon_s\rangle$ in Eq.~\eqref{eq:app_coupled_singlet_doublet} gives
\begin{align}
H_{\mathrm{low}}
={}& H_G - g_2 \int dx\,\cos(2\vartheta)
+ \frac{h}{2\pi}\int dx\,\partial_x\varphi + \cdots,
\label{eq:app_leading_bosonic_hamiltonian}
\end{align}
with
\begin{align}
g_2 &= -\frac{1}{\pi\alpha_0}
\left(m_d - \frac{g}{2}\langle\varepsilon_s\rangle\right), \notag\\
h &= -\sqrt{3}
\left(m_2 + \frac{g}{2}\langle\varepsilon_s\rangle\right).
\label{eq:app_effective_bosonic_couplings}
\end{align}
Thus, the interaction with the massive singlet renormalizes both the coefficient that tunes the doublet mass and the term responsible for shifting its characteristic momenta.

The leading continuum projection does not determine all operators generated at shorter length scales. We therefore supplement Eq.~\eqref{eq:app_leading_bosonic_hamiltonian} with the leading local vertices allowed by the microscopic symmetries. 
The microscopic internal symmetries (Eqs.~\eqref{eq:P} and \eqref{eq:K}) act on the angular field as
\begin{align}
\mathcal{P}:\ &\vartheta \rightarrow \vartheta + \pi,\ \
\mathcal{K}:\ \vartheta \rightarrow -\vartheta,\ \
\mathcal{PK}:\ \vartheta \rightarrow \pi - \vartheta.
\label{eq:app_angular_symmetries}
\end{align}
These transformations permit higher-harmonics $\cos(2n\vartheta)$ and exclude the corresponding sine terms. After the tuning term $\cos(2\vartheta)$, the leading allowed angular anisotropy is therefore $\cos(4\vartheta)$.

One-site translation cyclically permutes the three Majorana flavors, $\boldsymbol{\zeta} \rightarrow C\boldsymbol{\zeta}$, and similarly for $\boldsymbol{\eta}$. Projecting this transformation onto the doublet basis of Eq.~\eqref{eq:app_flavor_basis} gives
\begin{align}
\begin{pmatrix}
\zeta_1\\
\zeta_2
\end{pmatrix}
\rightarrow
\begin{pmatrix}
-\frac{1}{2} & \frac{\sqrt{3}}{2}\\
-\frac{\sqrt{3}}{2} & -\frac{1}{2}
\end{pmatrix}
\begin{pmatrix}
\zeta_1\\
\zeta_2
\end{pmatrix},
\end{align}
and the same transformation applies to $(\eta_1, \eta_2)^T$. Here, the lower indices $\mu = 1, 2$ label the two components of the real Majorana doublet. The transformation matrix represents a rotation by $-2\pi/3$ in the doublet plane. Since the chiral Majorana fields in Eq.~\eqref{eq:app_chiral_majoranas} are formed from $\eta_\mu$ and $\zeta_\mu$ without mixing the doublet indices, each chiral doublet $(\xi_r^1, \xi_r^2)^T$, with $r = L, R$, transforms by the same rotation. The chiral Dirac fields in Eq.~\eqref{eq:app_dirac_definition} are the two opposite-helicity combinations of these doublet Majoranas and therefore acquire the complex-conjugate phases
\begin{align}
\psi_L &\rightarrow e^{2\pi i/3}\psi_L,
&
\psi_R &\rightarrow e^{-2\pi i/3}\psi_R.
\label{eq:app_translation_dirac}
\end{align}
These phases are inherited from the commensurate lattice momenta $p_\pm = \pm Q$, with $Q = 2\pi/(3a_0)$, in Eq.~\eqref{eq:app_three_critical_momenta}. The interaction-induced momenta $q_\pm$ in Eq.~\eqref{eq:app_doublet_minima} instead describe the displacement of the doublet minima relative to these commensurate momenta. They do not modify the internal action of translation and are encoded in the bosonic theory by the term proportional to $\partial_x\varphi$.

Using the bosonization convention of Eq.~\eqref{eq:app_bosonization_dictionary}, the transformation in Eq.~\eqref{eq:app_translation_dirac} corresponds to
\begin{align}
\vartheta \rightarrow \vartheta,
\qquad
\varphi \rightarrow \varphi + \frac{4\pi}{3}.
\label{eq:app_translation_bosons}
\end{align}
Hence, $e^{in\varphi}$ is translation invariant only when $n$ is a multiple of three. The vertices $\cos\varphi$ and $\cos(2\varphi)$ are therefore forbidden, while $\cos(3\varphi)$ is the leading allowed dual vertex.

Including these symmetry-allowed perturbations, the effective Hamiltonian becomes
\begin{align}
H_{\mathrm{eff}} ={}& \frac{v}{2\pi}\int dx\left[K(\partial_x\vartheta)^2 + \frac{1}{4K}(\partial_x\varphi)^2 \right] \notag\\
&- \int dx\left[g_2\cos(2\vartheta) + g_4\cos(4\vartheta) + \ldots\right] \notag\\
&- \int dx \left[g_{3\varphi}\cos(3\varphi) + \ldots\right] \notag \\
&+ \frac{h}{2\pi}\int dx\,\partial_x\varphi.
\label{eq:app_effective_compact_boson}
\end{align}
Here, $g_4$ and $g_{3\varphi}$ are non-universal couplings included on symmetry grounds, while the ellipsis denotes higher symmetry-allowed harmonics. Equation~\eqref{eq:app_effective_compact_boson} reproduces the phenomenological Hamiltonian in Eq.~\eqref{eq:sine_gordon}.

In the convention of Eq.~\eqref{eq:app_effective_compact_boson}, vertex operators of charge $n$ has scaling dimensions
\begin{align}
\Delta\!\left[e^{in\vartheta}\right] = \frac{n^2}{4K},
\qquad
\Delta\!\left[e^{in\varphi}\right] = n^2K.
\label{eq:app_vertex_dimensions}
\end{align}
Thus, $\cos(2\vartheta)$ has scaling dimension $1/K$, $\cos(4\vartheta)$ has $4/K$, $\cos(3\varphi)$ has $9K$, and so on. 
Throughout the window
\begin{align}
\frac{1}{2} < K < 2,
\label{eq:app_K_window}
\end{align}
the tuning operator $\cos(2\vartheta)$ is relevant, whereas $\cos(4\vartheta)$, all higher angular harmonics, and $\cos(3n\varphi)$ with $n \geq 1$ are irrelevant.

The symmetry transformations of Eq.~\eqref{eq:app_angular_symmetries} also identify the leading continuum representatives of the two magnetizations, $m_x$ and $m_y$. 
Both $\cos\vartheta$ and $\sin\vartheta$ are odd under $\mathcal{P}$ and
invariant under translation, while $\mathcal{K}$ leaves $\cos\vartheta$ even and makes
$\sin\vartheta$ odd, precisely reproducing the transformation properties of $m_x$ and $m_y$.
The scaling dimensions fix what accompanies them. At the non-interacting critical point
$(\lambda, V_{nn}) = (1, 0)$, the $y$-magnetization vanishes at the $c = 3/2$ transition with
exponents $\beta = 3/8$ and $\nu = 1$~\cite{smacchia_pra_2011}, so that $\Delta[m_y] = \beta/\nu = 3/8$,
and $m_x$ must carry the same dimension. This is more than the doublet alone can supply. The
Jordan-Wigner strings carried by the microscopic magnetizations run over all three Majorana
flavors, so that the continuum representatives must have a contribution from every Majorana/Ising sector.
Since the three are identical at this point, each must contribute an operator of dimension
$1/8$, and the only fields of that dimension in a critical Ising theory are the order and
disorder fields~\cite{difrancesco_1997, mussardo_book_2020}.

The doublet contribution is then fixed by dimensional analysis and the symmetry transformations of Eq.~\eqref{eq:app_angular_symmetries}. 
Two of the three sectors form
the doublet, so together they must supply a dimension of $1/4$. At the non-interacting point $K = 1$,
Eq.~\eqref{eq:app_vertex_dimensions} gives $\Delta[e^{\pm i\vartheta}] = 1/4$,
matching the required value. Together with the symmetry assignments above, this identifies
$\cos\vartheta$ and $\sin\vartheta$ as the doublet contributions to $m_x$ and $m_y$, respectively.
The singlet contribution follows from parity. Since $\mathcal{P}$ is the total fermion parity,
it factorizes over the three sectors. Both $\cos\vartheta$ and $\sin\vartheta$ are already odd
under $\mathcal{P}$, and the magnetizations are themselves odd, so the remaining singlet field
must be even. Of the two dimension-$1/8$ fields of the singlet sector, only the disorder field
$\mu_s$ has this property. Therefore, up to non-universal amplitudes,
\begin{align}
m_x &\sim \langle\mu_s\cos\vartheta\rangle,
&
m_y &\sim \langle\mu_s\sin\vartheta\rangle.
\label{eq:app_order_parameters}
\end{align}
Throughout the $x$-FM and $y$-FM phases, and along the deconfined transition, the singlet
remains gapped in its disordered regime, and replacing $\mu_s$ by $\langle\mu_s\rangle$ leaves
$\cos\vartheta$ and $\sin\vartheta$ to determine the two magnetizations, with
$\Delta[m_x] = \Delta[m_y] = 1/(4K)$ as used in Sec.~\ref{sec:dqcl}.

Within the window of Eq.~\eqref{eq:app_K_window}, $g_2\cos(2\vartheta)$ is the only relevant cosine vertex. For $g_2 > 0$, it pins $\vartheta$ at $0$ or $\pi$, producing the $x$-FM phase, whereas $g_2 < 0$ selects $\vartheta = \pi/2$ or $3\pi/2$, producing the $y$-FM phase. Along $g_2 = 0$, all remaining symmetry-allowed cosine vertices are irrelevant, leaving a Gaussian theory with $c = 1$ and emergent $O(2)$ symmetry under $\vartheta \rightarrow \vartheta + \alpha$ and $\vartheta \rightarrow -\vartheta$. This is the DQC line separating the two ferromagnets in Sec.~\ref{sec:dqcl}. When $K$ decreases below $1/2$, $\cos(2\vartheta)$ also becomes irrelevant. For $2/9 < K < 1/2$, $\cos(3\varphi)$ remains irrelevant as well, allowing the floating phase discussed in Sec.~\ref{sec:strong_nn}.

Equation~\eqref{eq:app_order_parameters} also shows why pinning $\vartheta$ alone is insufficient to produce local magnetic order. In the present mass convention, $\langle\mu_s\rangle$ is non-zero only in the disordered Ising regime $m_s > 0$. On the opposite side of the singlet transition, where $m_s < 0$, i.e., below the $c = 1/2$ line in Figs.~\ref{fig:schem}(a) and \ref{fig:phase_nn}(a), the singlet sector is ordered and $\langle\mu_s\rangle = 0$. Both magnetizations therefore vanish even when $\vartheta$ is pinned. This accounts for the absence of local ferromagnetic order in the cluster and disordered phases shown in Fig.~\ref{fig:phase_nn}.

\subsection{Next-nearest-neighbor interactions and phase-diagram inversion}
\label{subsec:vnnn_inversion}

\begin{figure}[t]
\centering
\includegraphics[width=\linewidth]{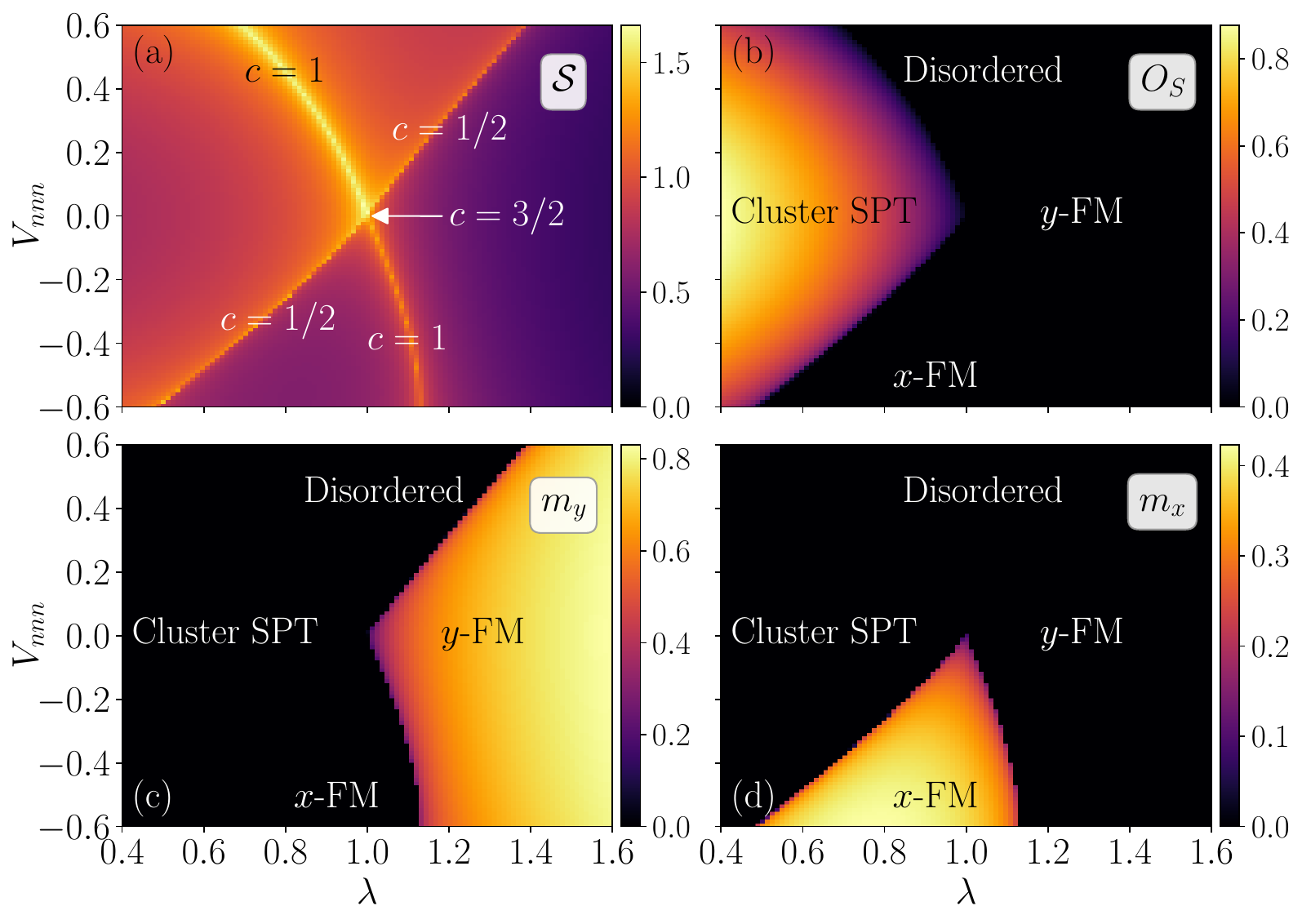}
\caption{(Color online.) Phase diagram of the CIM with the next-nearest-neighbor interaction in Eq.~\eqref{eq:app_nnn_interaction}, obtained using VUMPS with bond dimension $\chi = 128$. We show the bipartite entanglement entropy $\mathcal{S}$ (a), the string order parameter $O_S$ (b), and the magnetizations $m_y$ (c) and $m_x$ (d). For $V_{nnn} > 0$, the cluster and $y$-FM phases are separated by a disordered phase, whereas for $V_{nnn} < 0$, an intervening $x$-FM phase appears. The central charges indicated in (a) show that the $c = 1$ and $c = 1/2$ critical lines exchange their order across $V_{nnn} = 0$.}
\label{fig:phase_nnn}
\end{figure}

The same singlet-doublet analysis can be extended if we replace the nearest-neighbor interaction by a next-nearest-neighbor interaction,
\begin{align}
\hat{H}_{V}
= V_{nnn}\sum_j\hat{\sigma}^z_j\hat{\sigma}^z_{j + 2}.
\label{eq:app_nnn_interaction}
\end{align}
The nearest-neighbor interaction couples the three Majorana flavors in the cyclic order
$1 \rightarrow 2 \rightarrow 3 \rightarrow 1$, whereas the next-nearest-neighbor interaction couples them in the reverse order,
$1 \leftarrow 2 \leftarrow 3 \leftarrow 1$. Reordering the Majorana operators into the same reference convention introduces an additional minus sign in the contracted quadratic contribution. The resulting correction therefore has the same flavor structure, proportional to $C^2$, but enters the mass matrix with the opposite sign:
\begin{align}
M_{nnn}
= 2(\lambda - 1)\mathbb{I}_3
- \frac{4V_{nnn}}{\pi}C^2
+\mathcal{O}(V_{nnn}^2).
\label{eq:app_nnn_mass_matrix}
\end{align}
Consequently,
\begin{align}
m_s &= 2(\lambda - 1) - \frac{4V_{nnn}}{\pi}
+\mathcal{O}(V_{nnn}^2), \notag\\
m_d &= 2(\lambda - 1) + \frac{2V_{nnn}}{\pi}
+\mathcal{O}(V_{nnn}^2),
\label{eq:app_nnn_split_masses}
\end{align}
which are obtained from the nearest-neighbor results by the replacement
$V_{nn} \rightarrow -V_{nnn}$. The theory therefore predicts that the singlet and doublet critical lines exchange their order, and at a qualitative level produce a phase diagram reflected across the non-interacting axis.

Figure~\ref{fig:phase_nnn} confirms this prediction numerically. For $V_{nnn} > 0$, the string order vanishes before $m_y$ becomes non-zero, producing an intervening disordered phase bounded by $c = 1$ and $c = 1/2$ critical lines. For $V_{nnn} < 0$, a region with non-zero $m_x$, i.e., the $x$-FM phase, instead separates the cluster and $y$-FM phases, and the order of the two critical lines is reversed. The phase diagram is, therefore, the reflection of the nearest-neighbor result in Fig.~\ref{fig:phase_nn}, consistent with the replacement $V_{nn} \rightarrow -V_{nnn}$ predicted by the mass matrix.

\bibliography{draft.bbl}

\end{document}